\documentclass[11pt,a4paper,notoc]{article}
\pdfoutput=1
\usepackage{jcappub}
\title{Resolving astrophysical uncertainties in dark matter direct
  detection}
\author[a,b]{Mads T. Frandsen,}
\author[a]{Felix Kahlhoefer,}
\author[a,c]{Christopher McCabe,}
\author[a]{Subir Sarkar,}
\author[a,b]{and Kai Schmidt-Hoberg}
\affiliation[a]{Rudolf Peierls Centre for Theoretical Physics,
  University of Oxford, 1 Keble Road, Oxford OX1 3NP, United Kingdom}
\affiliation[b]{Theory Division, CERN, 1211 Geneva 23, Switzerland}
\affiliation[c]{Institute for Particle Physics Phenomenology, Durham
  University, South Road, Durham, DH1 3LE, United Kingdom}
\emailAdd{m.frandsen1@physics.ox.ac.uk}
\emailAdd{felix.kahlhoefer@physics.ox.ac.uk}
\emailAdd{christopher.mccabe@durham.ac.uk}
\emailAdd{s.sarkar@physics.ox.ac.uk}
\emailAdd{ksh@physics.ox.ac.uk}
\abstract{We study the impact of the assumed velocity distribution of
  galactic dark matter particles on the interpretation of results from
  nuclear recoil detectors. By converting experimental data to
  variables that make the astrophysical unknowns explicit, different
  experiments can be compared without implicit assumptions concerning
  the dark matter halo. We extend this framework to include the annual
  modulation signal, as well as multiple target elements. Recent
  results from DAMA, CoGeNT and CRESST-II can be brought into agreement if the velocity distribution is very anisotropic and thus allows a large modulation fraction.
  However constraints 
  from CDMS and XENON cannot be evaded by appealing to such astrophysical
  uncertainties alone.}
\keywords{Dark matter detectors, Dark matter experiments, Dark matter
  theory, Galaxy dynamics}
\arxivnumber{OUTP-11-52-P, CERN-PH-TH/2011-232, IPPP/11/65, DCPT/11/128}
\notoc
\begin{document}
\maketitle

\section{Introduction}

The direct detection of galactic dark matter (DM) particles in the
laboratory using shielded nuclear recoil detectors is among the most
challenging and high priority goals of astroparticle
physics. Presently several experiments are taking data with the
primary aim of securing a convincing detection and thereby confirming
the hypothesis that DM is in fact made of relic particles. Several
ambitious next-generation experiments are planned and from these we
hope to be able to extract the properties of the DM particle, most
notably its mass and cross-section for interactions with
nucleons. However, there are a number of uncertainties in the
interpretation of such experimental data. Foremost among these is our
poor knowledge of the distribution and dynamics of DM in our
Galaxy. In particular, the DM velocity distribution
$f(\boldsymbol{v})$ strongly affects the conclusions we can draw from
the data, for example whether the results of different experiments are
mutually consistent \cite{Kamionkowski:1997xg, Green:2002ht,
  Fairbairn:2008gz, MarchRussell:2008dy, Kuhlen:2009vh, Green:2010gw,
  Strigari:2009zb, McCabe:2010zh, Arina:2011si}.

Results from direct detection experiments are usually presented as a
signal or exclusion curve in the parameter plane of DM mass versus its
scattering cross-section, \emph{assuming} a Maxwell-Boltzmann (M-B)
velocity distribution with a cut-off at the escape velocity from the
Galaxy. This is also known as the Standard Halo Model (SHM). There are
more sophisticated dynamical models of halo DM but to study the
effects of these the plot has to be reproduced separately for all such
velocity distributions which can be cumbersome. 

One might wonder if
the experimental data can itself be used to infer the DM velocity
distribution. Unfortunately, this turns out to be rather difficult
\cite{Peter:2011eu, Drees:2007hr} because a nuclear recoil of energy
$E_\text{R}$ can originate from any DM particle which has a velocity
larger than some minimum velocity
$v_\text{min}(E_\text{R})$. Consequently, rather than probing the
velocity distribution, direct detection experiments actually measure
the velocity integral: $g(v_\text{min}) = \int_{v_\text{min}}
f(\boldsymbol{v})/v \, \text{d}^3v$.

Hence, it was suggested \cite{Fox:2010bz} that results from one
experiment be converted into $v_\text{min}$-space in order to predict
the event rate in a second experiment, without having to make any
assumptions concerning the astrophysics. While this approach works
well in comparing two experiments that are sensitive to a similar
range of DM velocities, it is difficult to compare experiments that
probe different regions of the velocity distribution. It would
therefore be preferable to have a framework in which the astrophysical
unknowns are made explicit so one can assess their impact on the
extraction of DM properties.

In this paper we investigate in detail the idea, also suggested in
Ref.\cite{Fox:2010bz}, of mapping \emph{all} experimental results into
$v_\text{min}$-space. To do so, we must know the properties of the DM
particles --- we will assume therefore that their mass can be inferred
either from collider experiments or the combined information from
several direct detection experiments (see
e.g. Ref.\cite{Peter:2011eu, Drees:2008bv}). In this sense, our approach is
complementary to the usual analysis of direct detection experiments,
wherein the astrophysical parameters are held fixed and the DM mass
and cross-section are allowed to vary.

Just as the usual presentation of results shows whether all
experiments are consistent for some range of DM mass and scattering
cross-section, our treatment will illustrate whether it is possible to
find a DM velocity distribution that can reconcile all such
experimental results. To demonstrate this, we focus on the DM mass
range 6--15 GeV motivated by recent claims of signals from DAMA,
CoGeNT and CRESST-II
\cite{Bernabei:2010mq,Aalseth:2011wp,Angloher:2011uu} and show that
$g(v_\text{min})$ can indeed be chosen so as to enable a consistent
description of these experiments. However, upper limits on signals
from XENON \cite{Aprile:2011hi} and CDMS \cite{Ahmed:2009zw,
  Ahmed:2010wy} \emph{cannot} be evaded by appealing to astrophysical
uncertainties alone. If all these experimental results are correct
then to resolve the tension between the signals and upper limits
requires non-standard DM interactions.

This paper is organised as follows: Section~\ref{sec:gofvmin}
introduces notation and sets out the analysis framework. In
Section~\ref{sec:measureconstrain} we discuss how a measurement of the
differential event rate in a direct detection experiment can be
translated into a measurement of $g(v_\text{min})$ and how an
experiment that does not observe a DM signal can constrain the
velocity integral. We present measurements and limits on
$g(v_\text{min})$ for the most relevant direct detection experiments
in Section~\ref{sec:results} and also assess the compatibility of
CoGeNT and CRESST-II. In Section~\ref{sec:modulations} we do a similar
analysis for experiments which observe an annual modulation of the
event rate and assess the compatibility of DAMA and CRESST-II. In
Section~\ref{sec:particle} we discuss briefly how varying the particle
physics properties of DM can affect the analysis, and present our
conclusions in Section~\ref{sec:conclusions}.

\section{The velocity integral}
\label{sec:gofvmin}

The differential event rate in the laboratory frame for the scattering
of DM particles on nuclei is given by
\begin{equation}
\frac{\text{d}R}{\text{d}E_\text{R}} = 
\frac{\rho\,\sigma_n}{2 m_\chi \mu_{n\chi}^2} 
C^2_\mathrm{T}(A,Z) F^2(E_\text{R}) g(v_\text{min}) \; ,
\label{eq:dRdE}
\end{equation}
where $\rho$ is the DM density, $m_\chi$ is the DM mass and
$\mu_{n\chi}$ is the reduced DM-nucleon mass. The `velocity integral'
$g(v_\text{min})$ is defined by
\begin{equation}
g (v_\text{min}, t) \equiv \int_{v_\text{min}}^\infty
\frac{f(\boldsymbol{v} + \boldsymbol{v}_\mathrm{E}(t))}{v}\,
\text{d}^3 v \; ,
\end{equation}
where $f(v)$ is the local DM velocity distribution evaluated in the
galactic rest frame, $v=|\boldsymbol{v}|$ and
$\boldsymbol{v}_\mathrm{E}(t)$ is the velocity of the Earth relative
to the galactic rest frame \cite{Gelmini:2000dm,Schoenrich:2009bx}.
The minimum velocity required for a DM particle to transfer an energy
$E_\text{R}$ to a nucleus is
\begin{equation}
v_\text{min}(E_\text{R}) = 
\sqrt{\frac{m_\mathrm{N} E_\text{R}}{2\mu^2}} \; ,
\end{equation}
where $m_\mathrm{N}$ is the mass of the target nucleus and $\mu$ the
corresponding reduced mass of the DM-nucleus system.

To avoid clutter in our formulae, we define $C_\mathrm{T}(A, Z) \equiv
\left(f_p/f_n Z + (A-Z)\right)$, where $A$ and $Z$ are the mass and
charge numbers of the target nucleus and $f_{n,p}$ denote the
effective DM coupling to neutrons and protons, respectively. We assume
$f_n / f_p = 1$ unless explicitly stated otherwise.  Finally,
$\sigma_n$ is the DM-neutron cross-section at zero momentum transfer
and $F(E_\text{R})$ is the nuclear form factor, which encodes the loss
of coherence as the momentum transfer deviates from zero. We use the
Helm form factor from Ref.\cite{Lewin:1995rx}.

\begin{figure}[tb]
\begin{center}
{
\centering
\begin{minipage}[bt]{0.48\textwidth}
	\centering
\includegraphics[width=.95\columnwidth]{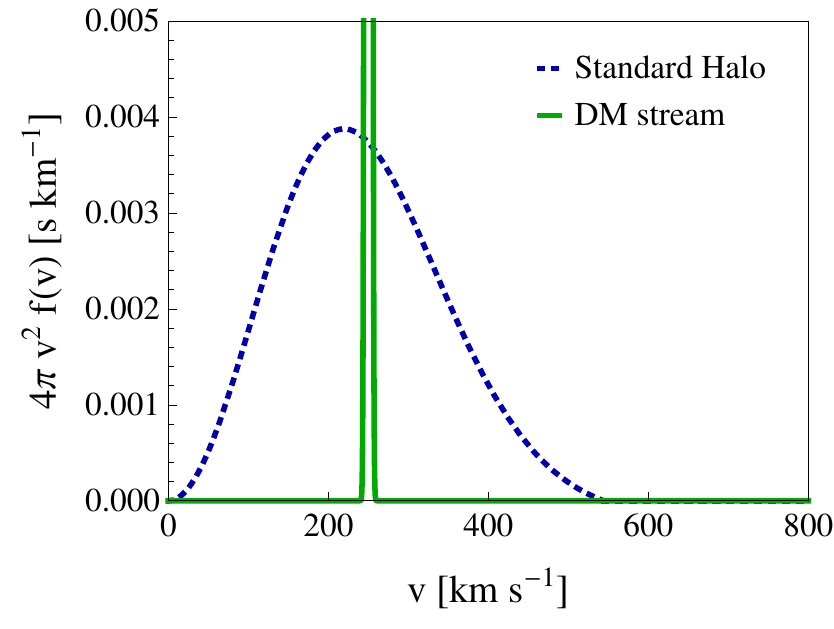}
\end{minipage}
\hfill
\begin{minipage}[bt]{0.48\textwidth}
	\centering
\includegraphics[width=.95\columnwidth]{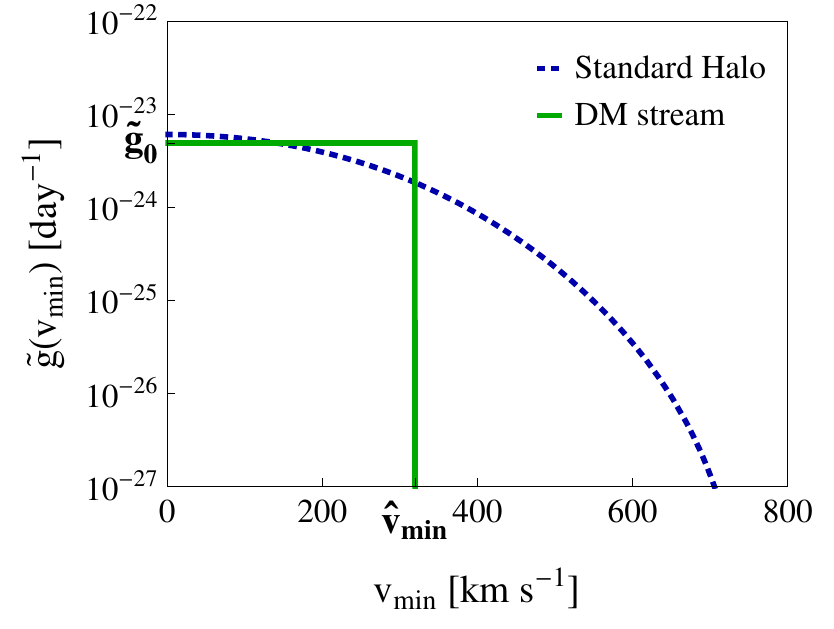}
\end{minipage} 
}
\caption{The one-dimensional velocity distribution $4 \pi v^2 f(v)$ (left) and the rescaled velocity integral $\tilde{g}(v_\text{min})$ (right) for the Standard Halo Model (blue, dotted) and a sharply peaked DM stream (green, solid). To calculate $\tilde{g}(v_\text{min})$, we have assumed $\sigma_n = 10^{-40}\ \text{cm}^2$, $\rho = 0.3 \ \text{GeV/cm}^3$ and $m_\chi = 15$ GeV. The velocity of the DM stream is $\boldsymbol{v}_\text{s} = (250,\, 0,\, 0)$ km/s, giving $\hat{v}_\text{min} = |\boldsymbol{v}_\text{s} - \boldsymbol{v}_\text{E}| \approx 330$ km/s.}
\label{fig:testfunction}
\end{center}
\end {figure}

For our purposes, it will be convenient to absorb the DM mass,
cross-section and density into the definition of the velocity integral
and consider the rescaled velocity integral
\begin{equation}
\tilde{g}(v_\text{min}) = \frac{\rho \, \sigma_n}{m_\chi} \,
g(v_\text{min}) \; ,
\label{eq:rescaledgovmin}
\end{equation}
which has a typical value in the range 
$10^{-27}-10^{-22}\ \text{day}^{-1}$ (see
Figure~\ref{fig:testfunction}).

At this point, it is worth developing some intuition for
$\tilde{g}(v_\text{min})$ and elucidating its relation to the more
familiar DM velocity distribution $f(v)$. In the left panel of
Figure~\ref{fig:testfunction}, we show the canonical M-B velocity
distribution of the SHM, along with a sharply peaked function which could arise
from a stream of DM or alternatively can be thought of as a shifted
M-B distribution in the limit of small velocity dispersion. In the
right panel we show the corresponding $\tilde{g}(v_\text{min})$.\footnote{We neglect the time dependence of $g(v_\text{min})$ here, which would lead to a slight smearing of the step function.} As
observed earlier \cite{Fox:2010bu}, $\tilde{g}(v_\text{min})$ has the
important property that for \emph{any} velocity distribution it is a
decreasing function of $v_\text{min}$ since $f(v) \geq 0$. As the
velocity dispersion decreases, $\tilde{g}(v_\text{min})$ becomes more
like a step-function which is the minimal form consistent with this
property.

\section{Measurements and constraints}
\label{sec:measureconstrain}

We now discuss how information on $\tilde{g}(v_\text{min})$ can be
extracted from direct detection experiments.\footnote{A similar discussion is presented in the appendix of Ref.~\cite{Fox:2010bz}.} First we consider
experiments that observe a signal that can be interpreted in terms of
DM scattering, namely DAMA, CoGeNT and CRESST-II. In this case, the
measured signal can be directly translated into a measurement of
$\tilde{g}(v_\text{min})$. Then there are experiments that do not
observe a DM signal above their expected background and thus constrain the
differential DM scattering rate which converts into a bound on
$\tilde{g}(v_\text{min})$. Since $\tilde{g}(v_\text{min})$ may be
time-dependent in principle, the resulting limit is strictly speaking
on the \emph{average} of $\tilde{g}(v_\text{min})$ over the period the
corresponding experiment was taking data.

\subsection{Measuring the velocity integral}
\label{sec:measureg}

It is relatively straightforward to infer the value of
$\tilde{g}(v_\text{min})$ at $v_\text{min}(E_\text{R})$ from a nuclear
recoil detector such as CoGeNT that consists of a single element and
measures the differential event rate $\text{d}R/\text{d}E_\text{R}$ at
the recoil energy $E_\text{R}$. Inverting Eq.~\eqref{eq:dRdE}, we find
\begin{equation}
\tilde{g}(v_\text{min}) = 2 \mu_{n\chi}^2\,
\frac{1}{C^2_\mathrm{T}(A,Z) F^2(E_\text{R})}
\,\frac{\text{d}R}{\text{d}E_\text{R}} \; .
\end{equation}
For a real detector we must take the detector resolution $\Delta E$
and the detector efficiency $\epsilon(E_\text{R})$ into account. If
the detector measures the differential event rate in the interval
$\left[E_1, E_2\right]$, we can infer the value of
$\tilde{g}(v_\text{min})$ for $v_\text{min}$ in the interval
$\left[v_\text{min}(E_1 - \Delta E), v_\text{min}(E_2 + \Delta
  E)\right]$:
\begin{equation}
\tilde{g}(v_\text{min}) = 2 \mu_{n\chi}^2\,
\frac{1}{C^2_\mathrm{T}(A,Z) F^2(E_\text{R}) \epsilon(E_\text{R})} \,
\frac{\text{d}R}{\text{d}E_\text{R}} \; .
\label{eq:gtilde}
\end{equation}
The recoil energy where the nuclear form factor and the efficiency are
evaluated should in the range $E_1 < E_\text{R} < E_2$; we assume that
the energy bin is sufficiently small that these quantities do not vary
significantly with $E_\text{R}$ within the bin.\footnote{This
  assumption is good for CoGeNT and DAMA for the light DM that we
  consider. However it is not a good approximation for
  CRESST-II where $\epsilon(E_\text{R})$ can change rapidly within a
  single bin that contains the onset of a new detector module; to
  avoid this we bin the data so such regions are avoided (see
  Appendix~\ref{ap:experiments}).}

Additional care is required for experiments like DAMA and CRESST-II
which have targets with more than one element. The DAMA detector has
both sodium and iodine nuclei, the latter being much heavier than the
former ($m_{\text{I}}\approx 5.5 m_{\text{Na}}$). For the light DM we
consider, the kinetic energy of the recoiling iodine nucleus is below
the energy threshold of the detector so it is a good approximation to
assume that all contributions to $\text{d}R / \text{d}E_\text{R}$ come
from scattering on Na nuclei and use Eq.~\eqref{eq:gtilde} taking the
mass fraction of Na in the detector into account (see
Section~\ref{sec:modulations}).

For CRESST-II which contains oxygen, calcium and tungsten, the
situation is more complicated. Since tungsten is much heavier than
oxygen and calcium, we can ignore it as the corresponding recoil
energy is below the detector's energy threshold. However oxygen and
calcium are relatively close in mass so extracting information on
$\tilde{g}(v_\text{min})$ becomes more complicated. We discuss these
difficulties and how to deal with them in Appendix~\ref{ap:isotopes}.

\subsection{Constraining the velocity integral}

\label{sec:constrain}

Next we consider experiments that do not measure the differential
event rate but are rather able to bound $\text{d}R/\text{d}E_\text{R}$
over some energy range. This is usually converted into a bound on
$\sigma_n$ for a given DM mass, assuming a galactic halo model such as
the SHM. Instead we wish to use the bound on $\text{d}R /
\text{d}E_\text{R}$ to constrain $\tilde{g}(v_\text{min})$ for
different values of $v_{\text{min}}$.\footnote{It is straightforward
  to put a bound on $\tilde{g}(v_\text{min})$ from experiments that
  use a single element; difficulties arising in experiments with more
  than one type of nuclei are discussed in
  Appendix~\ref{ap:isotopes}.}

Since $f(v)\geq0$, the following inequality holds for any value of
$v_\text{min}$:
\begin{equation}
\label{eq:inequal}
\tilde{g}(v_\text{min}) \geq \tilde{g}(\hat{v}_\text{min})\,
\Theta(\hat{v}_\text{min} - v_\text{min})\, ,
\end{equation}
where $\tilde{g}(\hat{v}_\text{min})$ is a constant and $\Theta(x)$ is
the Heaviside step function. Of
all velocity integrals that have $\tilde{g}(\hat{v}_\text{min}) =
\tilde{g}_0$, the one defined by \mbox{$\tilde{g}(v_\text{min}) =
  \tilde{g}_0\, \Theta(\hat{v}_\text{min} -v_\text{min})$} thus
predicts the smallest number of events in any given experiment.  Of
course \mbox{$g(v_\text{min}) \propto \Theta(\hat{v}_\text{min}
  -v_\text{min})$} is not a realistic model for the galactic halo,
nevertheless it is a valid velocity integral that can be used to
predict the event rate for a given experiment. A more realistic halo
must satisfy Eq.~\eqref{eq:inequal} and will therefore necessarily
predict a \emph{larger} event rate. Consequently, if we can reject the
case when $\tilde{g}(v_\text{min}) = \tilde{g}_0\,
\Theta(\hat{v}_\text{min} - v_\text{min})$, we can also reject any
other other halo model giving $\tilde{g}(\hat{v}_\text{min}) =
\tilde{g}_0$.

\begin{figure}[tb]
\begin{center}
{
\centering
\includegraphics[width=.50\columnwidth]{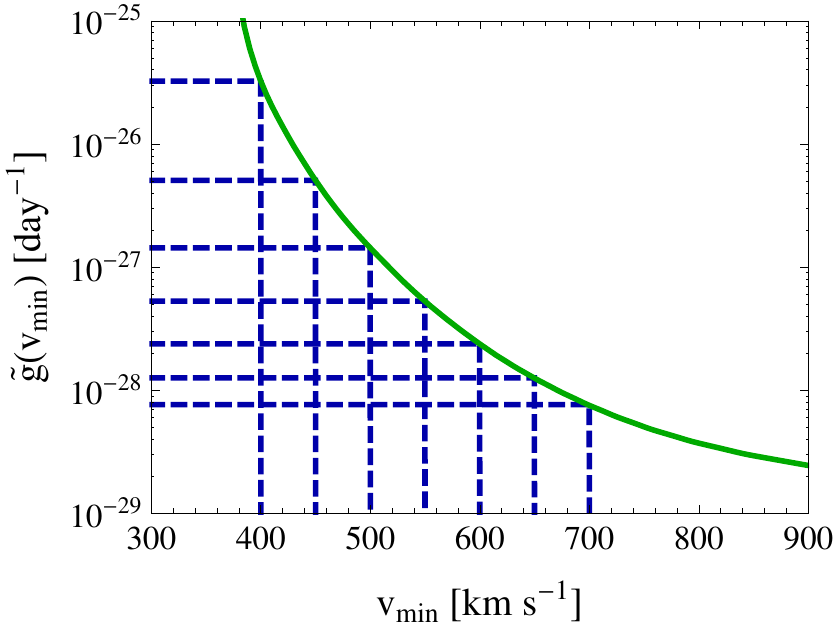}
}
\caption{An illustration of the method developed in Section~\ref{sec:constrain}. The blue dashed lines correspond to velocity integrals of the form $\tilde{g}_0\, \Theta(\hat{v}_\text{min} -v_\text{min})$ for different choices of $\tilde{g}_0$ and $\hat{v}_\text{min}$. The predicted event rate in the XENON100 detector for each of these velocity integrals can be rejected by the data from XENON100. The green line shows the resulting bound on $\tilde{g}(v_\text{min})$.}
\label{fig:illustration}
\end{center}
\end {figure}

Hence if an experiment places an upper bound on $\text{d}R /
\text{d}E_\text{R}$, we can correspondingly bound
$\tilde{g}(v_\text{min})$ by fixing $v_\text{min}=\hat{v}_\text{min}$
and finding the smallest value of $\tilde{g}(\hat{v}_\text{min})$ such
that the predicted event rate for \mbox{$\tilde{g}(v_\text{min}) =
  \tilde{g}(\hat{v}_\text{min})\, \Theta(\hat{v}_\text{min}
  -v_\text{min})$} is \emph{larger} than the measured value (at a
given confidence level). Repeating this procedure for all
$\hat{v}_\text{min}$, we obtain a continuous upper bound on
$\tilde{g}(\hat{v}_\text{min})$ --- see
Figure~\ref{fig:illustration}. If the exclusion curve thus obtained
touches (or crosses) the $\tilde{g}(\hat{v}_\text{min})$ curve for a
particular $f(v)$, this model will be excluded by the experimental
data at the same confidence level used to construct the exclusion
bound. In other words, if the bound on $\tilde{g}(v_\text{min})$ lies
below (some of) the values of $\tilde{g}(v_\text{min})$ implied by the
measured recoil spectrum at DAMA, CoGeNT or CRESST-II, it will not be
possible to find a halo model that consistently describes all
measurements and evades all experimental bounds.

Note that the converse of this statement is not necessarily true: even
if $\tilde{g}(v_\text{min})$ stays below a given exclusion curve for
all values of $v_\text{min}$, this does not imply that the model is
\emph{not} excluded by the corresponding experiment. By construction,
we have ensured only that the predicted differential event rate in
each individual energy bin is below the respective experimental
limit. However, if the measured velocity integral remains close to the
exclusion limit over a wide range of $v_\text{min}$, the predicted
total event rate, i.e.\ the integral of the differential event rate
over several bins, may be excluded at the chosen confidence level.

\section{Results}
\label{sec:results}

\begin{figure}[!t]
\centering
\includegraphics[width=\textwidth,clip,trim=5 15 5 20]{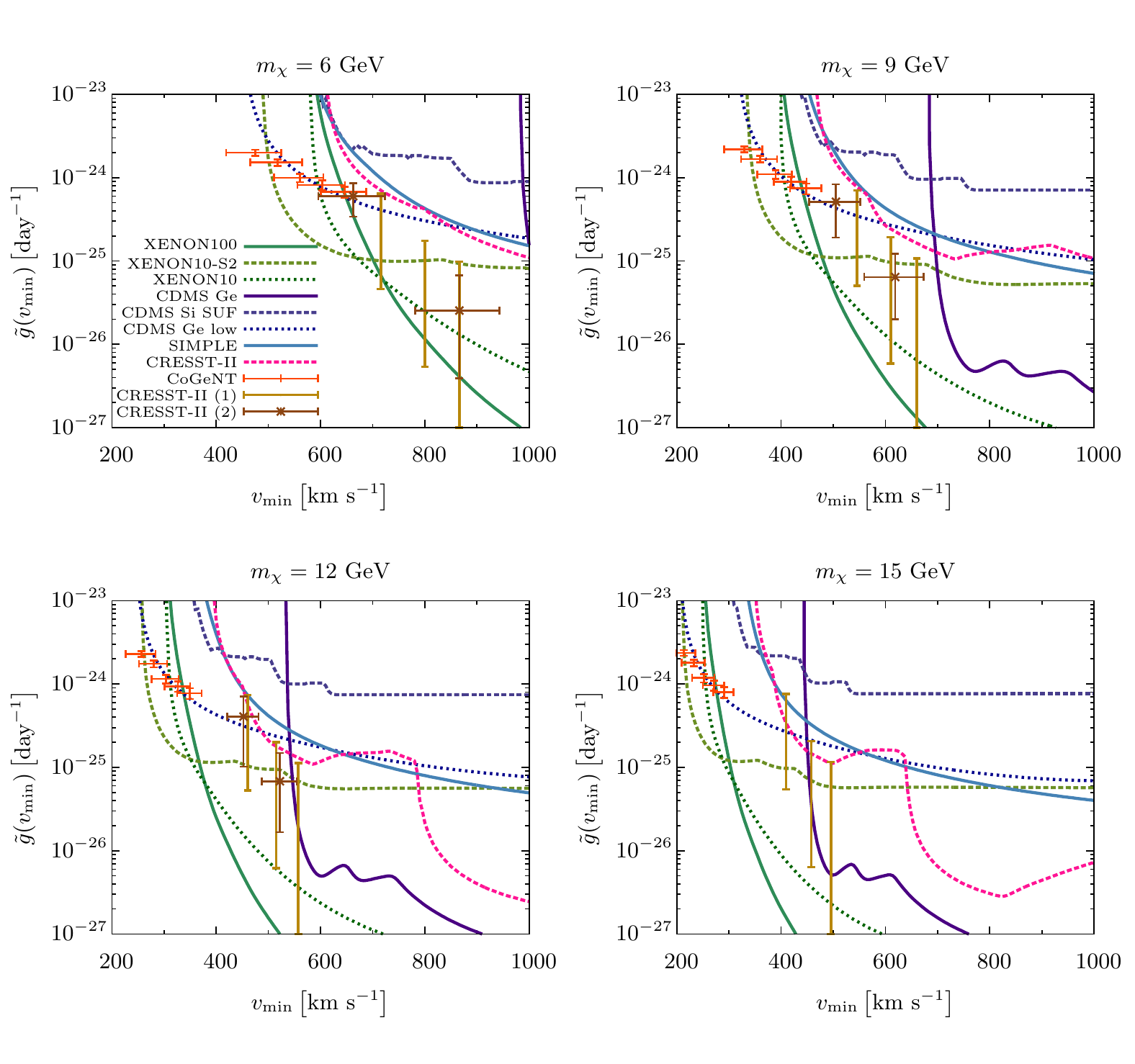} 
\caption{Measurements and exclusion bounds of the velocity integral
  $\tilde{g}(v_\text{min})$ for different DM masses $m_\chi$. The
  DM interaction is spin-independent and elastic with $f_n /
  f_p = 1$. Values of $\tilde{g}(v_\text{min})$ above the lines are
  excluded with at least $90\%$ confidence. For $m_\chi = 6$, 9 and 12 GeV the data points for CRESST-II have been obtained using two different methods, as described in Appendix~\ref{ap:isotopes}. It is not possible to find any
  model for the DM halo that provides a consistent description of all experiments.}
\label{fig:gofvmin}
\end {figure}

Now we use the framework developed in the previous section to analyse
recent results from direct detection experiments (discussed in
Appendix~\ref{ap:experiments}). We use the events reported by CoGeNT
and CRESST-II to measure $\tilde{g}(v_\text{min})$ and the null
results from XENON, CDMS, SIMPLE and the CRESST-II commissioning run
to constrain it.  Our results are shown in Figure~\ref{fig:gofvmin}
for various values of the DM mass $m_\chi$.  As we increase $m_\chi$
the exclusion curves as well as the CoGeNT data points move according
to $v_\text{min}\sim 1/m_\chi$ (which holds when $m_\chi <
m_\mathrm{N}$) --- the CRESST-II data points and exclusion curves
scale differently because the mass of oxygen is comparable to $m_\chi$
in the lower panels. It should be clear from this consideration that the CoGeNT and CRESST-II data points are always excluded for $6 \, \text{GeV} < m_\chi < 15 \, \text{GeV}$.

For all masses considered, one can see the obvious conflict between
the measurements by CoGeNT and CRESST-II and the exclusion bounds from
other experiments. In each case, most of the measured points lie above
one (or more) of the exclusion curves. We conclude that a consistent
description of all experiments is not possible for \emph{any} model of
the DM halo if the DM particles undergo elastic spin-independent
scattering with $f_n / f_p = 1$. There is no functional form for
$\tilde{g}(v_\text{min})$ that would allow a DM interpretation of the
CoGeNT or CRESST-II data consistent with other experiments.

\subsection{A consistent description of CoGeNT and CRESST-II}
\label{sec:CandC}

Even though we cannot find a halo model that provides a consistent
description of \emph{all} experiments, it is seen from
Figure~\ref{fig:gofvmin} that CRESST-II and CoGeNT probe
$\tilde{g}(v_\text{min})$ at different ranges of $v_\text{min}$.
Therefore it should be possible to choose $\tilde{g}(v_\text{min})$
such that we obtain a consistent description for these two
experiments. This choice must be different from the SHM for which the
best-fit DM regions of CRESST-II and CoGeNT do \emph{not} overlap
\cite{Angloher:2011uu, Kelso:2011gd}.

\begin{figure}[tb]
\begin{center}
{
\includegraphics[width=\textwidth,clip,trim=5 10 5 5]{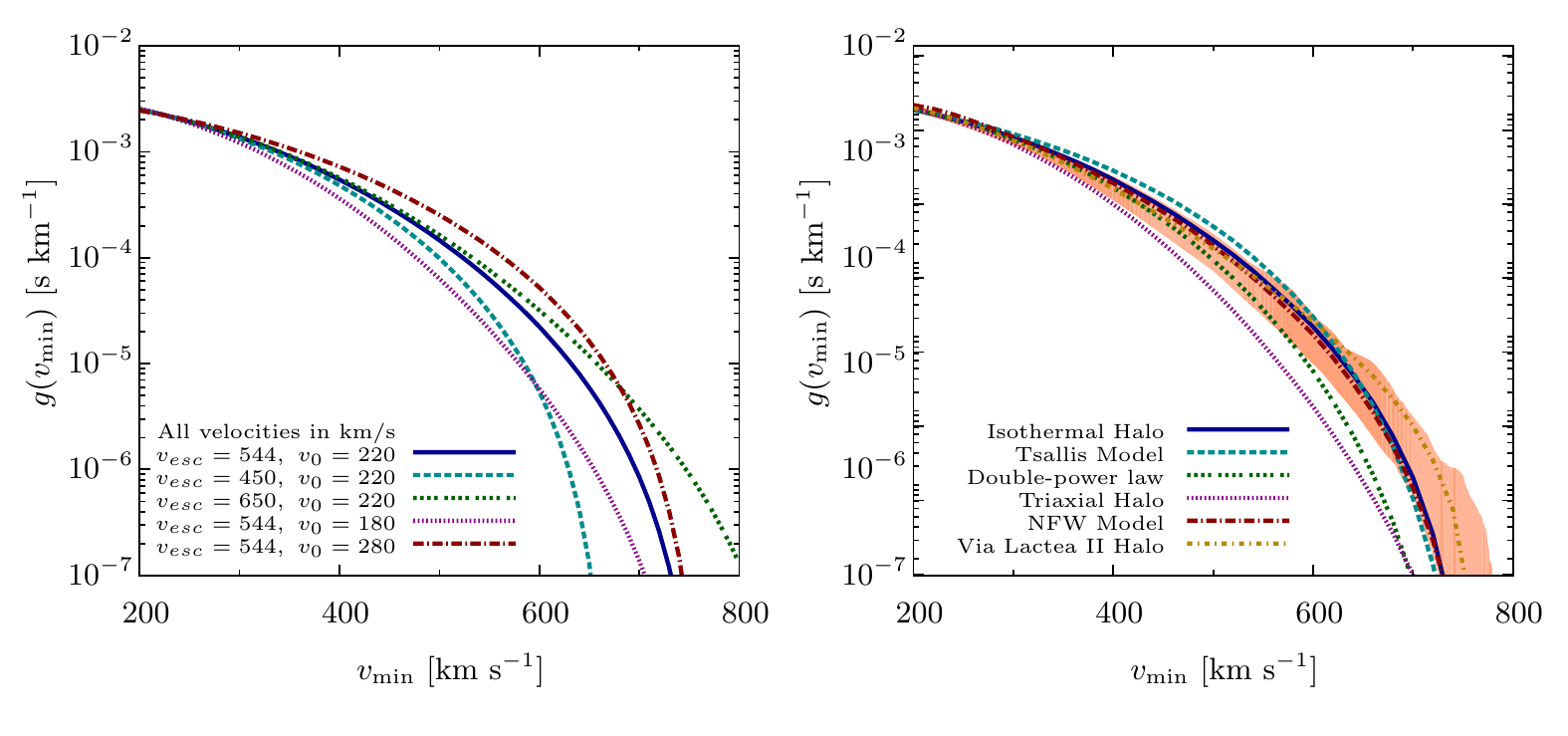} 
}
\caption{The velocity integral $g(v_\text{min})$ for the SHM with different parameters (left) and alternative descriptions of the DM halo introduced in Appendix~\ref{ap:halos} (right). The shaded region corresponds to the values of $g(v_\text{min})$ observed in the GHALO$_{\text{s}}$ simulation \cite{Kuhlen:2009vh}.}
\label{fig:HaloModels}
\end{center}
\end {figure}

Of course, we cannot vary $\tilde{g}(v_\text{min})$ arbitrarily~--- in
the end, the velocity integral must arise from a reasonable
self-consistent model of the DM halo. Therefore in Figure~\ref{fig:HaloModels}, we examine the range of predictions for $\tilde{g}(v_\text{min})$ from a variety of reasonable models of the galactic halo. First we note that
$\tilde{g}(v_\text{min})$ can change considerably even in the context
of the SHM, if we vary $v_0$ and $v_\text{esc}$ within their
observational bounds (see the left panel of
Figure~\ref{fig:HaloModels}). 
Secondly, the SHM is unlikely to be an accurate description; indeed many alternative models and parameterisations for the halo exist in the literature. We present an overview in
Appendix~\ref{ap:halos} and show the corresponding velocity integrals
in the right panel of Figure~\ref{fig:HaloModels}. Even for a fixed
choice of $v_\text{esc}$ and $v_0$ there is a large spread in the
predictions of the velocity integral, especially close to the cut-off.

\begin{figure}[tb]
\centering
\includegraphics[width=0.5\textwidth,clip,trim=5 10 0 15]{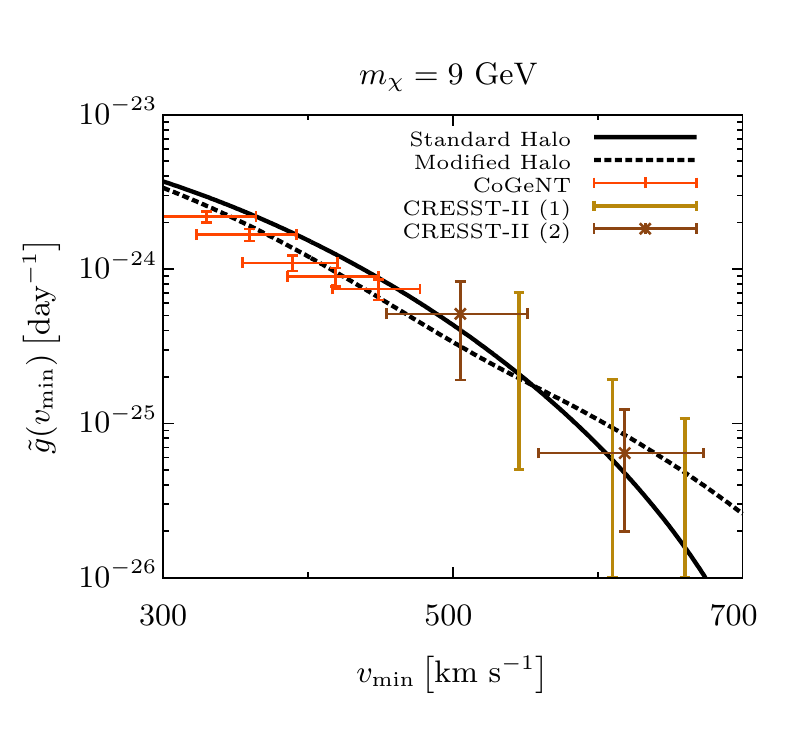} 
\caption{Measurements of the velocity integral $g(v_\text{min})$ compared to the predictions of the SHM and the Modified Halo Model. In both cases, we have assumed $\sigma_n = 10^{-40} \, \text{cm}^2$ and $\rho = 0.3 \, \text{GeV/cm}^3$. The Modified Halo Model provides a consistent description of the data from CoGeNT and CRESST-II for a DM particle with $m_\chi=9$ GeV.}
\label{fig:SHM}
\end {figure}

\begin{figure}[tb]
\begin{center}
\begin{minipage}[bt]{0.48\textwidth}
	\centering
\includegraphics[width=.95\columnwidth]{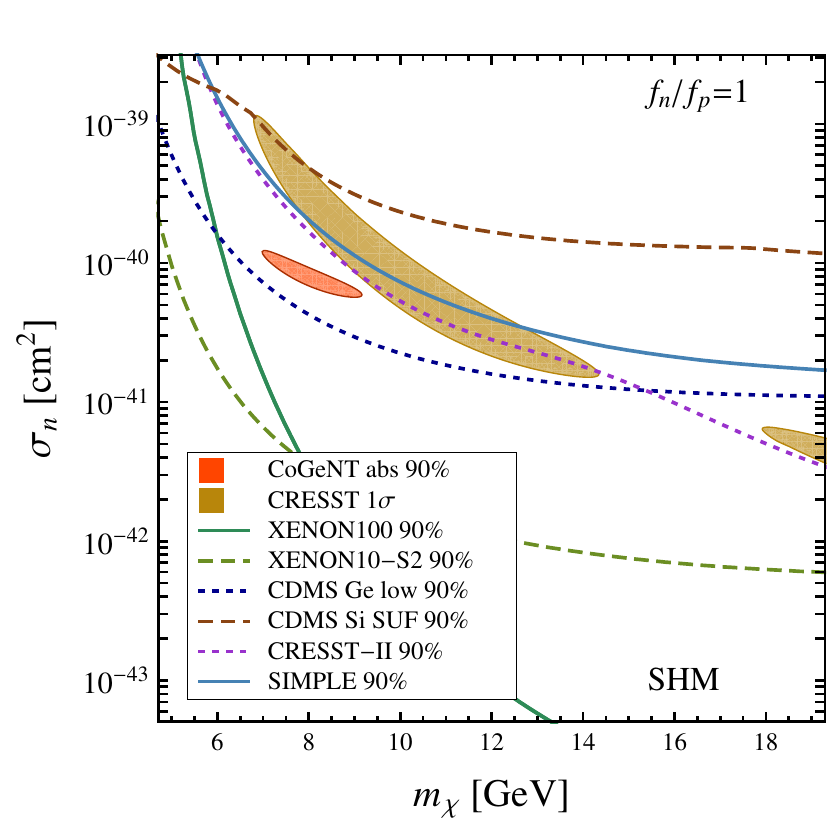}
\end{minipage}
\hfill
\begin{minipage}[bt]{0.48\textwidth}
	\centering
\includegraphics[width=.95\columnwidth]{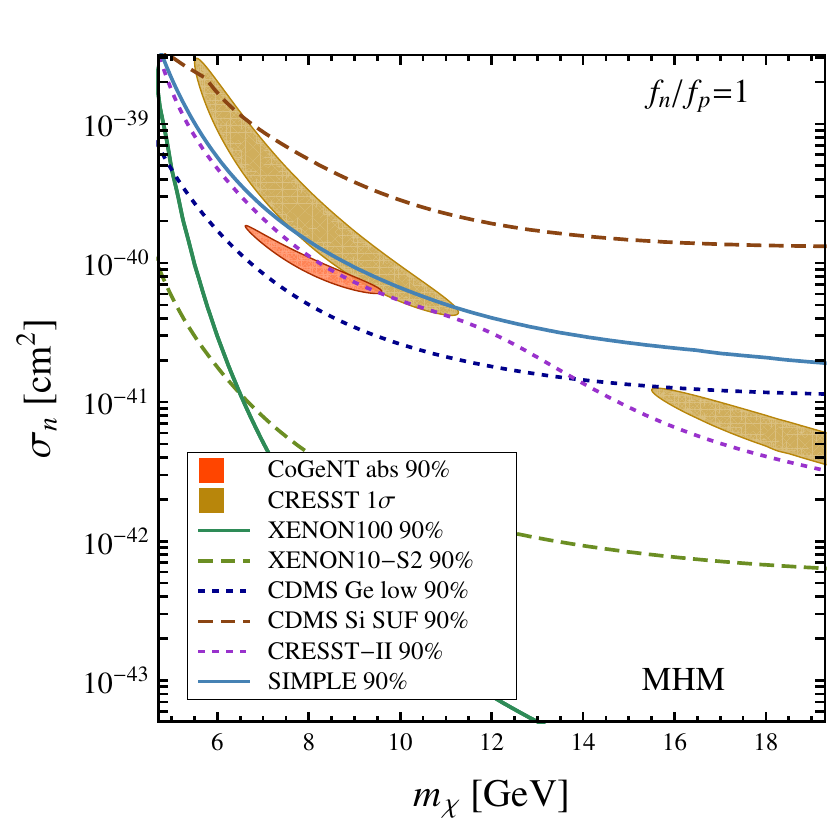}
\end{minipage} 
\caption{$90\%$ confidence regions inferred from the absolute rate observed by CoGeNT and $1\sigma$ regions from CRESST-II as well as the $90\%$ exclusion limits from other experiments for the SHM (left) and the Modified Halo Model (right). By construction, the Modified Halo Model leads to an overlap of the CoGeNT and CRESST-II regions.}
\label{fig:traditional}
\end{center}
\end{figure}

We conclude that given the spread in the predictions for
$\tilde{g}(v_\text{min})$ it should be possible to bring CoGeNT and
CRESST-II into better agreement. In this context it is instructive to
ask why CRESST-II favours larger DM masses than CoGeNT (see left panel
of Figure~\ref{fig:traditional}). To understand this, we show the SHM
prediction for $\tilde{g}(v_\text{min})$ in Figure~\ref{fig:SHM}
together with the measurements from CoGeNT and CRESST-II for $m_\chi =
9$ GeV. We observe that the SHM prediction of the rescaled velocity
integral below 500 km/s is slightly too flat to fit the CoGeNT data,
thus favouring smaller DM masses. Moreover there is an additional
constraint (which is not apparent in Figure~\ref{fig:SHM}) from the
fact that CoGeNT does not observe a signal at higher energies. This
constraint additionally disfavours the case $m_\chi \geq 9$ GeV if the
rescaled velocity integral is too flat. Consequently if we want to
push the CoGeNT region to larger DM mass, we need to make the rescaled
velocity integral steeper below 500 km/s. As a result our new velocity
integral will predict fewer events from oxygen in the lowest bins of
CRESST-II because $\tilde{g}(v_\text{min})$ becomes smaller around 500
km/s. To compensate, we must increase the contribution from calcium in
the lowest bins, corresponding to larger values of
$\tilde{g}(v_\text{min})$ around 600 km/s (see also the discussion in
Appendix~\ref{ap:isotopes}).

We can easily achieve this goal by adding two SHM-like contributions
--- one with relatively small velocity dispersion and low cut-off
which dominates at low values of $v_\text{min}$, and a second one with
much larger $v_0$ and $v_\text{esc}$. Of course, the resulting
velocity distribution is \emph{not} a self-consistent model of the DM
halo; it is only a convenient parameterisation to illustrate how the
velocity integral can be changed to simultaneously accommodate several
direct detection experiments. Nevertheless, it is not inconceivable
that such a velocity integral can be realised in models with a strong
anisotropy, where the tangential component with low velocity
dispersion dominates at low energies and the radial component with
high dispersion dominates at high energies. We shall refer to this new
velocity integral as the Modified Halo Model (MHM).

We show the corresponding best-fit regions and exclusion limits in the
$m_\chi$ -- $\sigma_n$ plane in Figure~\ref{fig:traditional}. As
desired, the CoGeNT and CRESST-II region\footnote{For CRESST-II we
  show the $1\sigma$ best-fit region using the publicly available data
  as stated in Appendix~\ref{ap:experiments}. Note that this
  corresponds roughly to the $2\sigma$ region published by the CRESST
  collaboration \cite{Angloher:2011uu}.} now overlap at $m_\chi \sim
9$~GeV. The common parameter region is, nevertheless, clearly excluded
by XENON and CDMS, an observation which we could equally well have
made from Figure~\ref{fig:gofvmin}.

\section{Measurements and constraints for modulation amplitudes}

\label{sec:modulations}

We have seen that for all the cases considered, there is strong
tension between the exclusion limits from null results and the event
rate observed by CoGeNT and CRESST-II. This seems to suggest that at
least some of the events seen by CoGeNT and CRESST-II may not be due
to DM scattering but a new source of background~\cite{CollarTAUP}. We
will therefore now consider a more specific signature of DM
interactions, namely the annual modulation of the signal due to the
motion of the Earth relative to the galactic rest frame. This will
help to identify the DM signal if the background is not
time-dependent.

In fact, two experiments have observed an annual modulation of their
signal, namely DAMA \cite{Bernabei:2010mq} and CoGeNT
\cite{Aalseth:2011wp}. By considering the modulation in $v_\text{min}$
space, it was observed \cite{McCabe:2011sr} that the two modulation
signals are (marginally) compatible with each other, independent of
astrophysics. Now we wish to include these annual modulations into our
analysis of $g(v_\text{min})$ to determine whether these signals are
compatible with the null results from other experiments. We first
discuss these constraints without making any assumptions about the modulation fraction
and then discuss how the modulation fraction can
itself be reasonably constrained in order to obtain stronger
experimental bounds.

The time dependence of $g(v_\text{min}, t)$ can  be approximately
parameterised as
\begin{equation}
g(v_\text{min}, t) = g(v_\text{min})\left[1+A(v_\text{min})\cdot
  \cos\left(2\pi\frac{t -
    t_0(v_\text{min})}{{1\ \text{yr}}}\right)\right] \;,
\end{equation}
where $g(v_\text{min})$ is the time average of
$g(v_\text{min},t)$. Note that as emphasised earlier
\cite{Kuhlen:2009vh}, \emph{both} the modulation fraction $A$ and the
peak date $t_0$ can depend in general on $v_\text{min}$. For future
use we define the modulation amplitude $\Delta g(v_\text{min})$ as
\begin{align}
\label{eq:deltag}
\Delta g(v_\text{min}) & \equiv \frac{1}{2}\left[g(v_\text{min},
  t_0(v_\text{min})) - g(v_\text{min}, t_0(v_\text{min}) +
  0.5\ \text{yr})\right] = A(v_\text{min}) g(v_\text{min}) ,
\end{align}
and introduce the rescaled modulation amplitude: $\Delta
\tilde{g}(v_\text{min}) = \Delta g(v_\text{min}) \sigma_n \rho/m_\chi$.

\begin{figure}[!tb]
\begin{center}
\includegraphics[width=\textwidth,clip,trim=5 20 5 15]{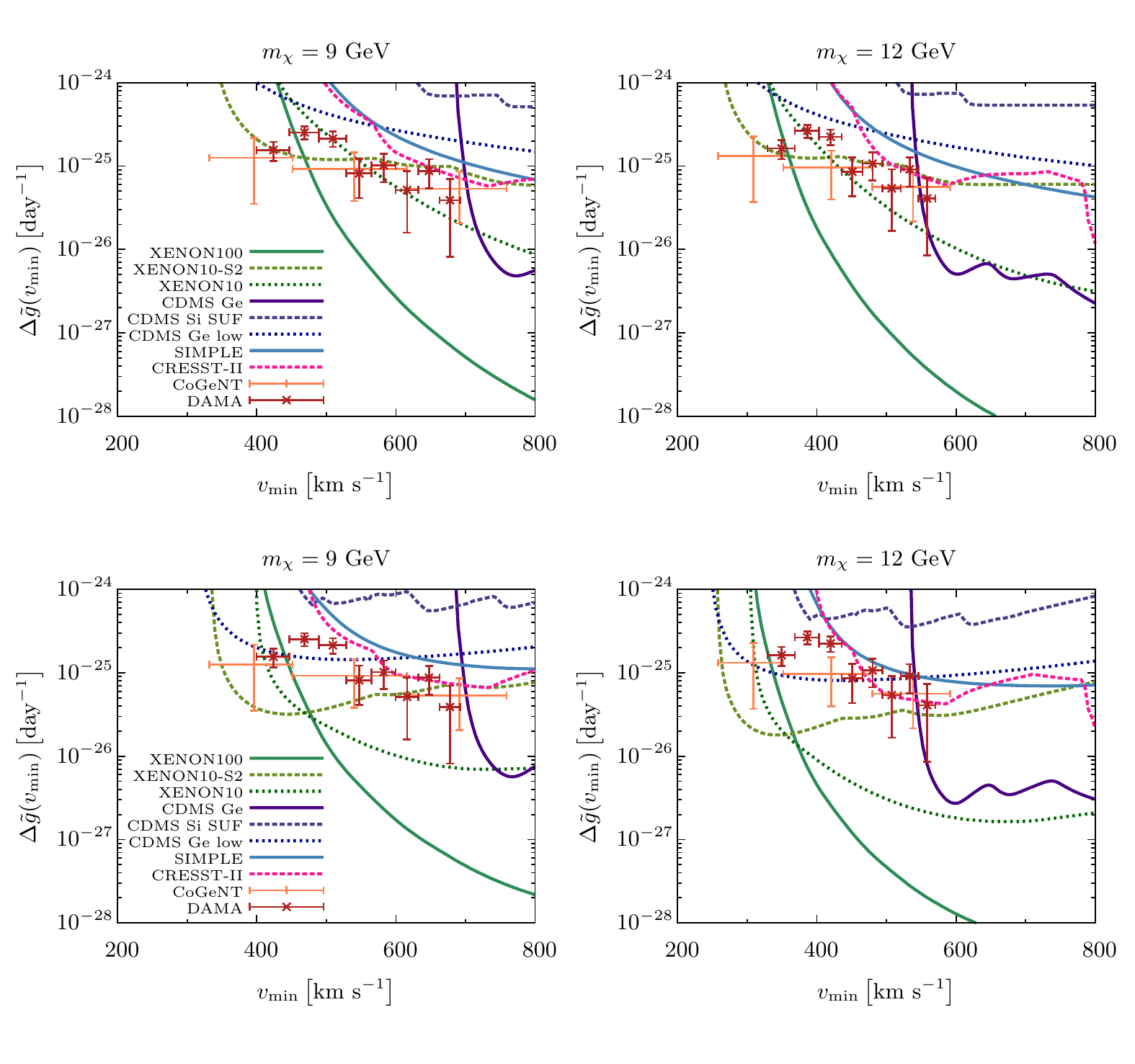} 
\caption{Measured values of $\Delta \tilde{g}(v_\text{min})$ from DAMA and
  CoGeNT compared to the exclusion limits from other experiments. For the upper panels, \emph{no} assumptions on the modulation fraction have been made, for the lower panels, we assume that the modulation fraction is bounded by the red line in the right panel of Figure~\ref{fig:ModulationConstraints}. Even for weak assumptions on the modulation fraction, there is significant tension between the different experiments, most notably it is impossible to find a DM velocity distribution that describes the observed modulations and evades the bound from XENON100.}
\label{fig:modplot100}
\end{center}
\end{figure}

First we need to constrain the function $t_0(v_\text{min})$. CoGeNT
and DAMA probe largely the same region of $v_\text{min}$ space, so it
would be inconsistent to assume different phases for the two
experiments. It was shown \cite{McCabe:2011sr} that the phase measured
by DAMA and by CoGeNT are consistent at the $90\%$ confidence level
for sufficiently low DM mass. We will therefore take
$t_0(v_\text{min}) = 146$ days which is the best-fit value from DAMA.\footnote{Varying $t_0$ within the $1\sigma$ region allowed by DAMA changes the best-fit value for the CoGeNT modulation amplitude by about $20\%$, which does not significantly affect any of our conclusions.}
As the modulation fraction satisfies $A \leq 1$, we see from
Eq.~\eqref{eq:deltag} that \mbox{$\Delta \tilde{g}(v_\text{min}) \leq
  \tilde{g}(v_\text{min})$}. We can easily check whether DAMA and
CoGeNT satisfy this inequality by plotting the respective measurements
of $\Delta \tilde{g}(v_\text{min})$ on top of the constraints for
$\tilde{g}(v_\text{min})$ from other experiments which we obtained
above. In complete analogy to Section~\ref{sec:measureg} we can
calculate $\Delta \tilde{g}(v_\text{min})$ from the modulation of the
event rate, $\Delta\text{d}R/\text{d}E_\text{R}$, seen by DAMA and
CoGeNT
\begin{equation}
\Delta \tilde{g}(v_\text{min}) = 2 \mu_{n\chi}^2 \,
\frac{1}{C^2_\mathrm{T}(A,Z) F^2(E_\text{R})}
\,\Delta\frac{\text{d}R}{\text{d}E_\text{R}} \; .
\end{equation}
The resulting values of $\Delta \tilde{g}(v_\text{min})$ are shown in
Figure~\ref{fig:modplot100}. We observe that the bounds from the
XENON10 and XENON100 experiments remain strong even if we do not make
any assumptions about the modulation fraction. However other
experiments do not significantly constrain $\Delta
\tilde{g}(v_\text{min})$. We consider therefore whether it is
reasonable to make stronger assumptions about the modulation fraction
and thus obtain more stringent experimental bounds.

\subsection{Constraining the modulation fraction}

\begin{figure}[tb]
\begin{center}
\includegraphics[width=\textwidth,clip,trim=5 10 5 5]{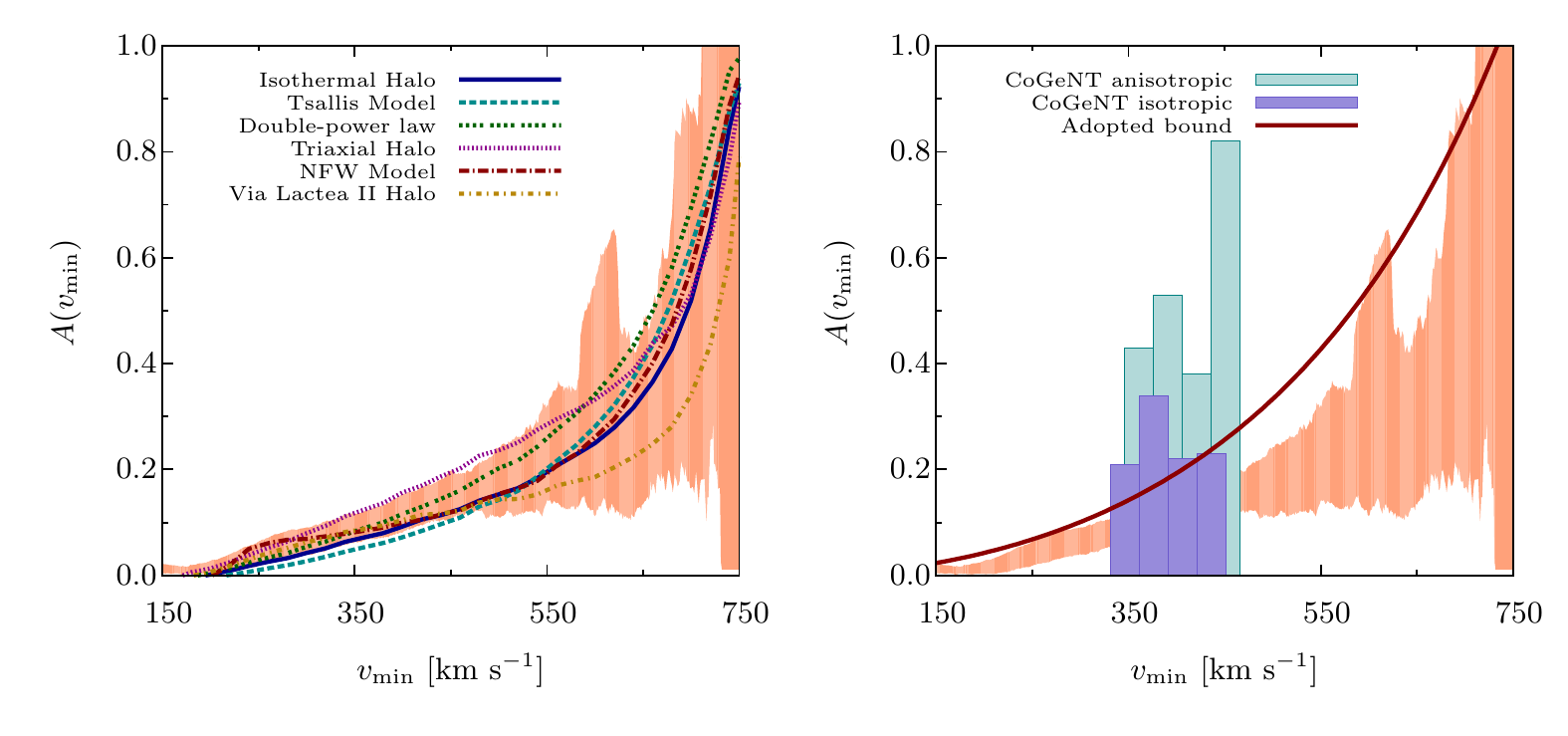} 
\caption{Left: The velocity integral and the modulation fraction for different descriptions of the galactic DM halo. Right: The constraint for the modulation fraction extracted from the CoGeNT data (boxes) together with the bound that we will adopt (red solid line). The shaded region in both panels corresponds to the modulation fractions  observed in the GHALO$_{\text{s}}$ simulation \cite{Kuhlen:2009vh}.}
\label{fig:ModulationConstraints}
\end{center}
\end {figure}

We will now discuss what can be reasonably assumed about the
modulation fraction given known models of the galactic halo, and how
it can be constrained once the velocity integral has been
measured. The predicted modulation fraction for various halo models
are shown in the left panel of
Figure~\ref{fig:ModulationConstraints}. We observe that for most
values of $v_\text{min}$ it is significantly below 100\%. Note that a
modulation fraction of 100\% implies that no signal is observed at $t_0 + 0.5$ yr, which is possible only if $v_\text{min} > v_\text{esc} + v_\text{E}(t_0 + 0.5\ \text{yr})$.

The empirical bound that we will adopt is shown as the red line in the right
panel of Figure~\ref{fig:ModulationConstraints}. Note that this is a
weak bound in the sense that we allow modulation fractions much larger
than the value expected in any reasonable description of the DM
halo. If we require that the modulation fraction satisfies
$A(v_\text{min}) \leq A^\text{max}(v_\text{min})$, any direct
detection experiment that constrains $\tilde{g}(v_\text{min}) \leq
\tilde{g}^\text{max}(v_\text{min})$ will also constrain $\Delta
\tilde{g}(v_\text{min})$ according to:
\begin{equation}
\Delta \tilde{g}(v_\text{min}) = A(v_\text{min})
\tilde{g}(v_\text{min}) \leq A^\text{max}(v_\text{min})
\tilde{g}^\text{max}(v_\text{min})\; .
\end{equation}
The resulting measurements and constraints for $\Delta
\tilde{g}(v_\text{min})$ are shown in the lower panels of Figure~\ref{fig:modplot100}. Even
with our weak assumption concerning the modulation fraction, the
bounds on $\Delta\tilde{g}(v_\text{min})$ have become significantly
stronger. Decreasing the mass of the DM particle now helps
significantly because at high values of $v_\text{min}$ larger
modulation fractions are allowed so the exclusion limits become less
stringent. However for small masses it becomes increasingly difficult
to explain the modulation observed by CoGeNT above 2 keVee because
these recoil energies correspond to very large velocities, where we
expect $\tilde{g}(v_\text{min})$ and therefore
$\Delta\tilde{g}(v_\text{min})$ to be small.

We now consider whether the modulation fraction can be constrained further
using measurements of $\tilde{g}(v_\text{min})$. Comparing the left
panel of Figure~\ref{fig:ModulationConstraints} to the right panel of
Figure~\ref{fig:HaloModels}, we observe that there is a direct
correspondence between the velocity integral and the modulation
fraction: the steeper the slope of the velocity integral, the larger
the modulation fraction. This result is intuitively clear --- if
$g(v_\text{min})$ changes rapidly with varying $v_\text{min}$ we can
also expect large changes between the rates observed in summer and
winter.  This can in fact be used to constrain the modulation fraction
once the velocity integral is known. We derive in
Appendix~\ref{ap:modfrac} that
\begin{equation}
A(v_\text{min}) = \frac{\tilde{g}(v_\text{min},\text{ summer}) -
  \tilde{g}(v_\text{min},\text{
    winter})}{\tilde{g}(v_\text{min},\text{ summer}) +
  \tilde{g}(v_\text{min},\text{ winter})} <
\frac{\tilde{g}(v_\text{min} - u) - \tilde{g}(v_\text{min} + u)}{2
  \tilde{g}(v_\text{min})} \; ,
  \label{eq:aconstraint}
\end{equation}
where $u = 29.8$ km/s if we assume an anisotropic halo and $u = 15.0$
km/s if we assume an isotropic halo.

As an example, we can use the CoGeNT data to derive a bound on the
modulation fraction for a given DM mass. For this purpose, we need to
bin the measured events in such a way that the resulting bin width in
$v_\text{min}$-space is equal to $u$ (see
Appendix~\ref{ap:experiments}). The resulting constraints are shown in
the right panel of Figure~\ref{fig:ModulationConstraints}. We observe
that assuming an isotropic halo, these constraints are actually quite
severe, limiting the modulation fraction to $20\%$ at
\mbox{$v_\text{min} = 400$ km/s}, which in fact coincides with our adopted bound for $A(v_\text{min})$. For an anisotropic halo, modulation
fractions of up to $40\%$ may be possible in the same range.

\begin{figure}[tb]
\centering
\includegraphics[width=0.5\textwidth,clip,trim=5 10 0 15]{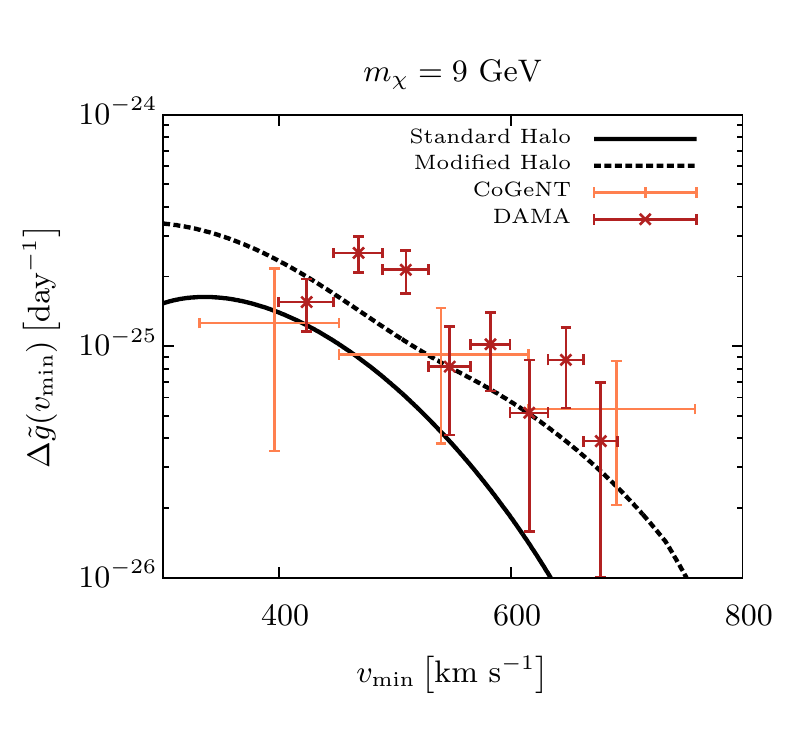} 
\caption{Measurements of the modulation amplitude $\Delta \tilde{g}(v_\text{min})$ compared to the predictions of the SHM and the Modified Halo Model. In both cases, we have assumed $\sigma_n = 10^{-40} \, \text{cm}^2$ and $\rho = 0.3 \ \text{GeV/cm}^3$. While in the SHM, the predicted modulation amplitude for these parameters is too small to explain the observed modulations, we can get a sufficiently large modulation amplitude in the Modified Halo Model.}
\label{fig:SHMmod}
\end {figure}

This observation is also relevant for a self-consistent DM
interpretation of the CoGeNT experiment. To bring the modulation
observed by CoGeNT into agreement with the measured event rate, a
relatively large modulation fraction is
required~\cite{Frandsen:2011ts}. If CoGeNT were to reject a
significant part of its signal as background, as suggested in
Ref.\cite{CollarTAUP}, even larger modulation fractions will be
necessary for self-consistency. The constraint on the modulation
fraction derived above can then be used to conclude that the DM halo
must be anisotropic.

\subsection{A consistent description of DAMA and CRESST-II}

As can be seen from Figure~\ref{fig:modplot100}, DAMA and CoGeNT probe the same region of
$v_\text{min}$ space, so it is not possible to improve their agreement
by changing the halo model as we did for CoGeNT and
CRESST-II. Fortunately, such a modification is not necessary because
DAMA and CoGeNT favour roughly the same modulation amplitude. The more
interesting question is whether this modulation amplitude is
compatible with the average value of $\tilde{g}(v_\text{min})$
inferred from CRESST-II and the CoGeNT unmodulated event rate.

\begin{figure}[tb]
\begin{center}
\begin{minipage}[bt]{0.48\textwidth}
	\centering
\includegraphics[width=.95\columnwidth]{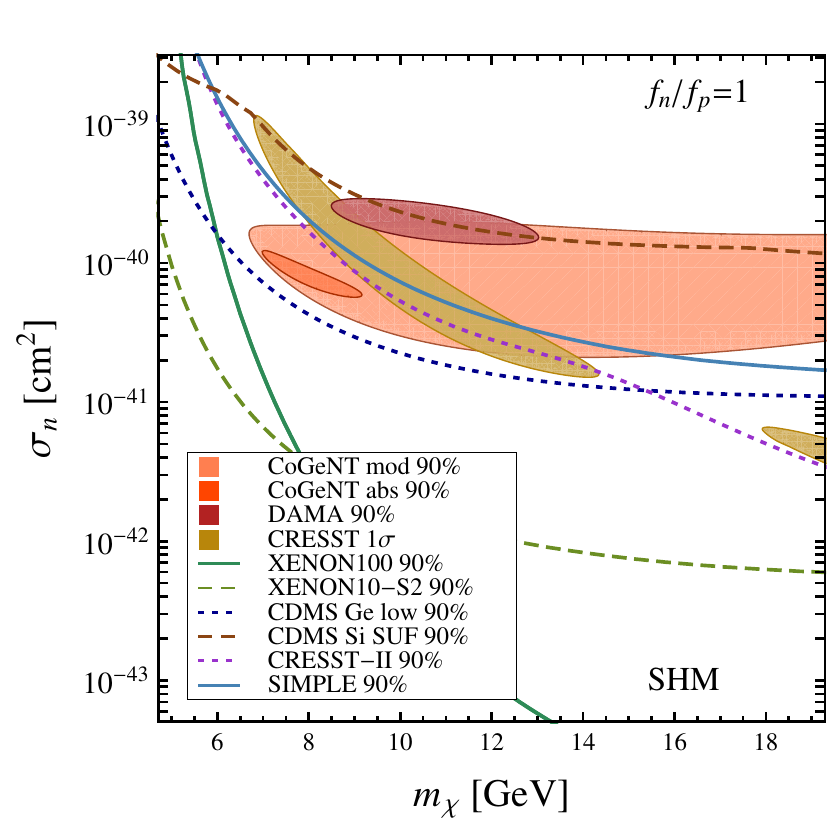}
\end{minipage}
\hfill
\begin{minipage}[bt]{0.48\textwidth}
	\centering
\includegraphics[width=.95\columnwidth]{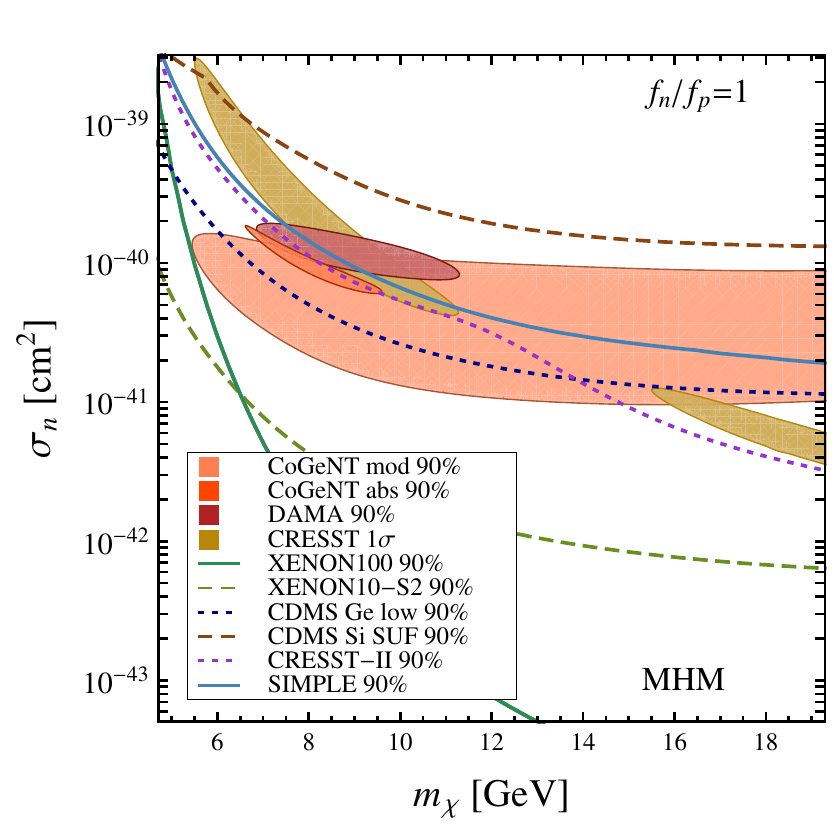}
\end{minipage} 
\caption{$90\%$ confidence regions from CoGeNT (inferred from both the modulation and the absolute rate) as well as the $1\sigma$ region from CRESST-II and the DAMA region for the SHM (left) and the Modified Halo Model (right). The $90\%$ exclusion limits from other experiments are also shown. By construction of the Modified Halo Model, the four best-fit regions are now in good agreement.}
\label{fig:traditionalDAMA}
\end{center}
\end{figure}

For a cross-section of $\sigma_n \approx 10^{-40}\, \text{cm}^2$, the
SHM prediction of the modulation amplitude is too small to account for
the modulation seen by DAMA and CoGeNT
\cite{Frandsen:2011ts,Schwetz:2011xm,Farina:2011pw,Fox:2011px,McCabe:2011sr}.
This is obvious from plotting the SHM prediction for $\Delta
\tilde{g}(v_\text{min})$ together with the measured modulation
amplitudes (see Figure~\ref{fig:SHMmod}). However, we have already
noted in Section~\ref{sec:CandC} that the SHM must be modified if we
want to simultaneously explain CoGeNT and CRESST-II. Now we consider
if we can bring these experiments into agreement with DAMA (and
therefore also with the CoGeNT modulation) by additionally allowing a
larger modulation fraction.
Of course, for a given velocity integral the
modulation fraction must satisfy Eq.~\eqref{eq:aconstraint}, so we cannot choose arbitrarily large values.

For this purpose, we assume that $g(v_\text{min})$ is given by the
Modified Halo Model from Section~\ref{sec:CandC} and that the
modulation fraction saturates the bound $A^\text{max}(v_\text{min})$
shown in Figure~\ref{fig:ModulationConstraints}. 
By the reasoning in Appendix~\ref{ap:modfrac} these two
assumptions are compatible \emph{only} if the DM halo is highly
anisotropic. As is demonstrated in Figure~\ref{fig:SHMmod}, we then
obtain a sufficiently large modulation amplitude to describe the DAMA
and CoGeNT modulations. Consequently, the four best-fit regions in the
traditional $\sigma_{n}-m_\chi$ parameter plane are now in good
agreement (Figure~\ref{fig:traditionalDAMA}). However these are of
course excluded by XENON and CDMS. Because of this obvious
contradiction we wish to emphasise again that we do not consider
present data sufficient to actually determine the velocity integral or
the modulation amplitude.  We use it only to illustrate how our method
can be used to bring future, more reliable, data sets into agreement.

\subsection{Additional contributions to the local dark matter density}

Our discussion of $\tilde{g}(v_{min})$ is quite independent of the
origin of the local DM density. However to predict the velocity
integral (Figure~\ref{fig:HaloModels}) and the modulation amplitude
(Figure~\ref{fig:ModulationConstraints}) we have assumed that the
local DM density is completely dominated by the contribution from the
galactic DM halo. In general there may be other significant
contributions to the local DM density, e.g.\ from DM `streams' and a
`dark disk'. We will now briefly discuss how these can alter the
theoretical predictions of $g(v_\text{min})$ and $A(v_\text{min})$.
 
\subsubsection*{Streams}

$N$-body simulations show that the DM velocity distribution is not a
smooth function but instead has a significant amount of
sub-structure~\cite{Vogelsberger:2008qb} due to the presence of tidal
streams. Such streams can have large velocities relative to the local
standard of rest and can therefore contribute to $g(v_\text{min})$ at
large values of $v_\text{min}$. The result is an edge in the velocity
integral (see Figure~\ref{fig:testfunction}) as well as a significant
increase in the modulation fraction~\cite{Savage:2006qr,
  Kelso:2011gd}. Nevertheless, the bound on the modulation fraction
derived in Appendix~\ref{ap:modfrac} remains valid. In other words, a
large modulation fraction is always visible as a steep decrease in the
velocity integral and can therefore be probed by measuring the
differential event rate (see also Ref.\cite{Natarajan:2011gz} for a
discussion on how to constrain DM streams with CoGeNT).

\begin{figure}[tb]
\begin{center}
\includegraphics[width=\textwidth,clip,trim=5 10 5 5]{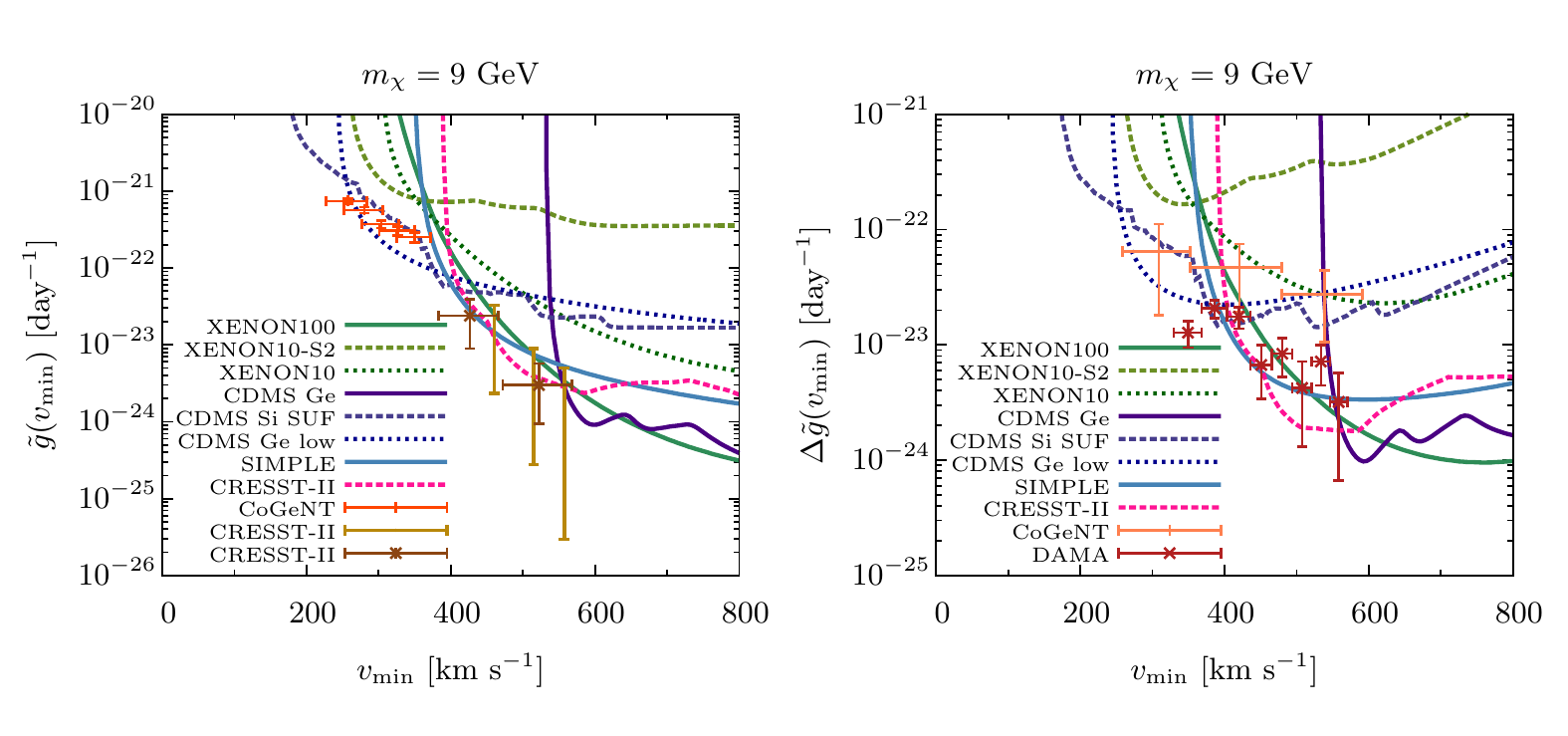} 
\caption{Measurements and exclusion bounds of the velocity integral   $\tilde{g}(v_\text{min})$ (left) and the modulation amplitude $\Delta \tilde{g}(v_\text{min})$ (right) under the assumption $f_n / f_p = -0.7$. As above, we have assumed that the modulation fraction is bounded by the red line in the right panel of Figures~\ref{fig:ModulationConstraints}. The bounds from liquid xenon experiments can mostly be evaded in this case but the bounds from CDMS-Si, SIMPLE, and the CRESST-II commissioning run become much more important.}
\label{fig:fnfp}
\end{center}
\end {figure}

\subsubsection*{Dark disk}

Most $N$-body simulations of the galactic DM halo neglect the effect
of baryons. Baryonic matter is expected to effect the formation of a
DM disk that can contribute a similar amount to the local DM density
as the halo~\cite{Ling:2009eh,Read:2009iv}. Such a dark disk is
expected to have a much smaller velocity dispersion than the galactic
halo and to be co-rotating with only a small lag. Consequently, a dark
disk will contribute to $\tilde{g}(v_\text{min})$ only for
$v_\text{min}\leq 250$ km/s --- a region that is irrelevant in the
context of light DM~\cite{MarchRussell:2008dy,
  Bruch:2008rx,Green:2010gw}. For DM particles with $m_\chi > 50$ GeV,
one can however expect a significant increase in both the velocity
integral and the modulation fraction at low energies.

\section{Varying particle physics}

\label{sec:particle}

All the results shown so far correspond to standard assumptions for
the interaction between DM particles and detector nuclei. Various
modifications of these assumptions have been proposed to reduce the
tension between various experiments. A particularly interesting
possibility is isospin-dependent DM couplings, i.e. $f_n \neq
f_p$. Choosing the ratio of couplings $f_n / f_p \simeq -0.7$ strongly
suppresses the bounds from all experiments using xenon
\cite{Kurylov:2003ra,Giuliani:2005my,Chang:2010yk,Feng:2011vu,Frandsen:2011ts,Schwetz:2011xm,Farina:2011pw,Fox:2011px,McCabe:2011sr},
which are otherwise the most constraining (see
Figure~\ref{fig:traditionalDAMA}). This observation has generated much
interest in DM models leading to such couplings
\cite{Feng:2011vu,DelNobile:2011je,Frandsen:2011cg,Cline:2011zr,Giuliani:2011py}.

\begin{figure}[tb]
\begin{center}
\begin{minipage}[bt]{0.48\textwidth}
	\centering
\includegraphics[width=.95\columnwidth]{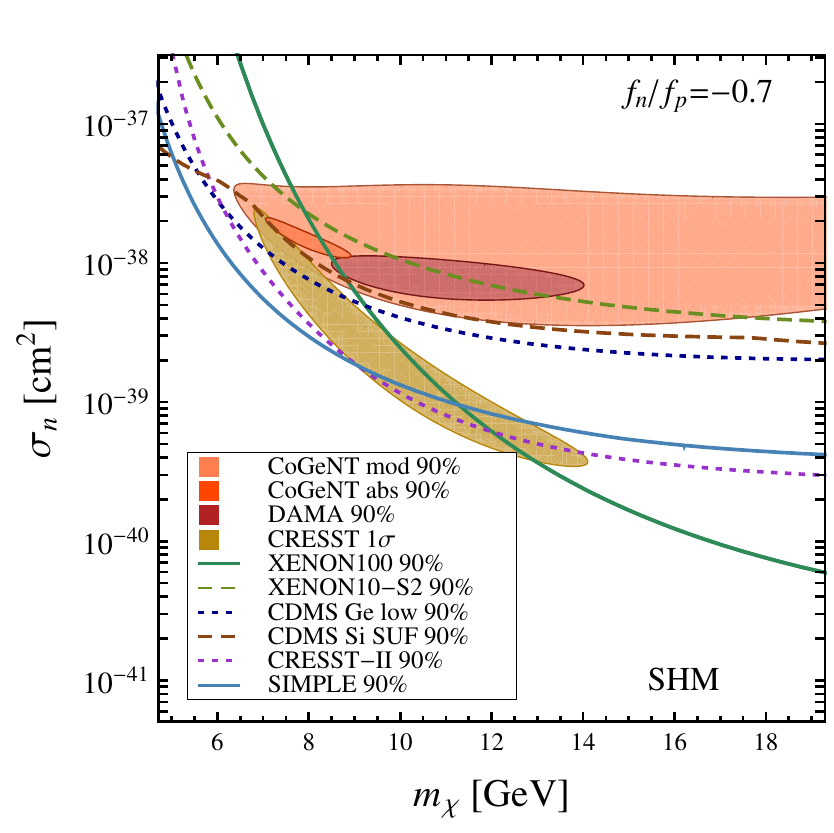}
\end{minipage}
\hfill
\begin{minipage}[bt]{0.48\textwidth}
	\centering
\includegraphics[width=.95\columnwidth]{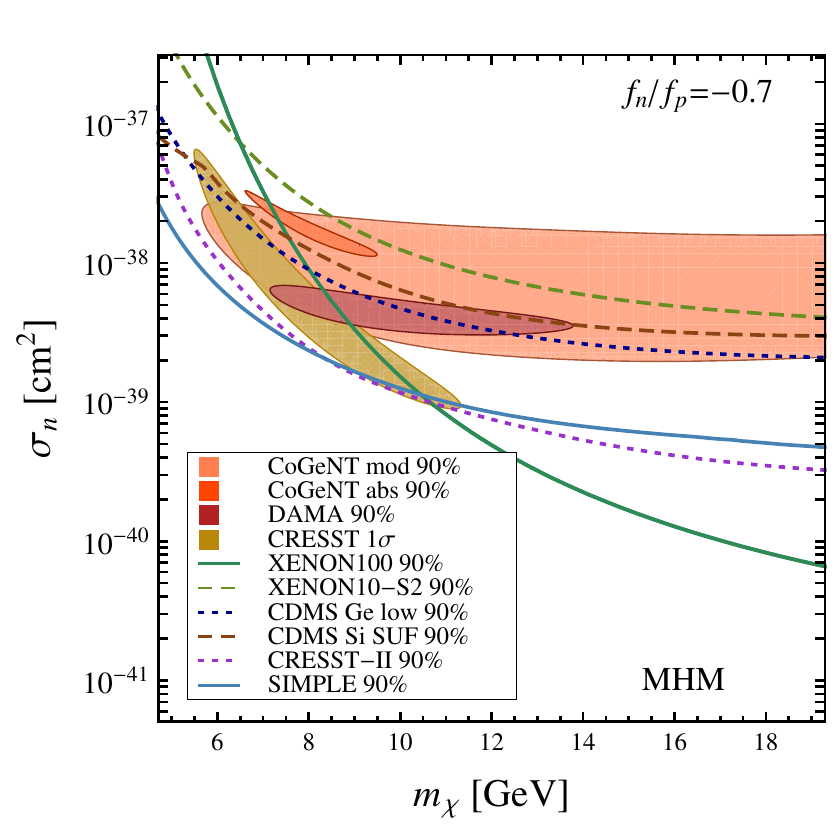}
\end{minipage} 
\caption{$90\%$ confidence regions from CoGeNT (inferred from both the modulation and the absolute rate) as well as the $1\sigma$ region from CRESST-II and the DAMA region for the SHM (left) and the Modified Halo Model (right) for the assumption $f_n / f_p = -0.7$. The $90\%$ exclusion limits from other experiments are also shown. For the MHM, the region of overlap from DAMA, CRESST-II and the CoGeNT modulation is only just excluded by current experimental limits.}
\label{fig:traditionalfnfp}
\end{center}
\end{figure}

We show the resulting measurements of $\tilde{g}(v_\text{min})$ and $\Delta \tilde{g}(v_\text{min})$ in Figure~\ref{fig:fnfp}. The bounds from XENON10 and CDMS-Ge can be evaded in this case but the bounds
from \mbox{CDMS-Si}, SIMPLE, and the CRESST-II commissioning run become more
important. Also, for $f_n / f_p \simeq -0.7$ the agreement between
CoGeNT and CRESST-II as well as the agreement between the DAMA and
CoGeNT modulations get considerably worse. Interestingly, the
agreement between CoGeNT and CRESST-II would improve considerably, if
the CoGeNT event rate were to be reduced through to the subtraction of
an additional background. The same observation can also be made from
Figure~\ref{fig:traditionalfnfp} (see also Refs.\cite{Kopp:2011yr,
  Kelso:2011gd}). For the MHM, the region of overlap from DAMA,
CRESST-II and the CoGeNT modulation is only just excluded by current
experimental limits.

\section{Conclusions}
\label{sec:conclusions}

We have discussed how measurements of the differential event rate in
nuclear recoil detectors can be used to extract information on the
velocity integral $g(v_\text{min})$. By converting experimental
results into $v_\text{min}$-space, we can compare the results from
various recent DM direct detection experiments without making any
assumptions concerning the properties of the DM halo. This strategy
has several direct applications.

The first is to compare experiments that observe a potential DM signal
with the null results from other direct detection experiments. To do
this we need to translate bounds on the differential event rate into
bounds on $\tilde{g}(v_\text{min})$. In order to exclude a certain
value $\tilde{g}(\hat{v}_\text{min})$ we must demonstrate that
\emph{any} velocity integral that takes this value necessarily
predicts an unacceptably large number of events in at least one of the
experiments. We have utilised this approach to demonstrate that
astrophysical uncertainties are \emph{not} sufficient to solve the
present conflicts between direct detection experiments. Specifically,
the values of $\tilde{g}(v_\text{min})$ favoured by CoGeNT and
CRESST-II are excluded by XENON and CDMS.

The second application is to assess the compatibility of several
experiments that observe a positive signal, independently of
astrophysical assumptions. If the experiments probe the \emph{same}
region of $v_\text{min}$ space, but measure contradictory values of
$\tilde{g}(v_\text{min})$, we can conclude that the two experiments
cannot be brought into agreement by changing astrophysics. However, if
the two experiments probe \emph{different} regions of $v_\text{min}$
space, either because they employ different targets or different
thresholds, it will often be possible to modify the velocity integral
in such a way that both experiments are brought into agreement~---
provided that a monotonically decreasing function fitting all data
exists.

We have demonstrated this possibility by considering the recent
signals seen by CoGeNT and CRESST-II. For this purpose, we have
developed a method to treat targets consisting of several different
nuclei. While the DM masses inferred from these measurements do not
agree for the SHM, we can modify the halo model to make the respective
best-fit regions overlap. To assess how reasonable such a modification
of the SHM is, we have studied a large variety of different dynamical
models for the galactic DM halo and calculated the respective
predictions for $\tilde{g}(v_\text{min})$. 
The corresponding spread in predictions for the velocity integral is rather large and justifies considering significant departures from the SHM value.

Similarly we can convert measurements of the modulation of the event rate into measurements of the modulation amplitude of the velocity integral $\Delta \tilde{g}(v_\text{min})$. Again such a conversion can be used to check whether several experiments observing an annual modulation are compatible and to confront them with constraints from null results. We find that the annual modulations observed by CoGeNT and DAMA are consistent with each other but in conflict with the bound from XENON100 even if we allow a modulation fraction of $100\%$. By constraining the modulation fraction with the absolute rate we have shown that such a large modulation fraction is in fact inconsistent with the CoGeNT data. Nevertheless for anisotropic halos it remains possible to have a sufficiently large modulation fraction to allow for a consistent description of CoGeNT, DAMA and CRESST-II, although the bounds from other experiments then become much stronger.

In conclusion, for standard interactions between DM particles and
nuclei, the tension between direct detection experiments cannot be
completely resolved by varying astrophysical parameters. If the
signals reported by DAMA, CoGeNT and CRESST-II are interpreted in
terms of DM while all null results are taken at face value,
non-standard interactions like isospin-dependent DM have to be
invoked. We have discussed this briefly and found that for $f_n/f_p =
-0.7$ experimental constraints are significantly weaker and can be
evaded by modifying the halo to have a larger modulation fraction. In
this case, the CoGeNT and DAMA modulations as well as the CRESST-II
event rate can be brought into agreement with all other experiments
except the bound from SIMPLE and the CRESST-II commissioning run. We
look forward to more data to help resolve the present puzzling
situation.

\section*{Acknowledgements}
We thank Wyn Evans, Matthew McCullough, John March-Russell and Andrew
Brown for useful discussions and Jens Schmaler for correspondence. FK
is supported by DAAD, KSH by ERC Advanced Grant (BSMOXFORD 228169) and
MTF and CM by STFC grants. MTF, KSH and SS thank the CERN theory group
for hospitality, especially during the TH-Institute DMUH'11 when this
project was initiated. We also acknowledge support from the UNILHC
network (PITN-GA-2009-237920) and an IPPP associateship for 2011-12
awarded to SS.
\begin{appendix}

\section*{Appendix}

\section{Including different target elements}

\label{ap:isotopes}

As demonstrated in Section~\ref{sec:measureg}, it is relatively easy
to convert energy spectra into $v_\text{min}$-space for a target
consisting only of a single element. However many experiments, for
example SIMPLE and CRESST-II, employ a combination of different
elements as target. In this case, a nuclear recoil of a given energy
corresponds to different $v_\text{min}$ depending on the recoiling
nucleus.

We discuss how to disentangle the arising ambiguities. Note that these
considerations may also be important for targets consisting of
different isotopes of a single element, especially when considering
isospin-dependent interactions.

\subsection{Constraining the velocity integral}

First, we discuss how an upper bound can be placed on the velocity
integral for a null result from an experiment consisting of more than
one element. The most obvious example is the SIMPLE experiment,
consisting of C$_2$ClF$_5$. We apply the method presented in
Section~\ref{sec:measureg} separately to each element (or isotope) by
requiring that the number of recoil events expected for this element
alone does not significantly exceed the total number of events
observed.

\begin{figure}[tb]
\centering
\includegraphics[width=0.5\textwidth,clip,trim=5 10 0 15]{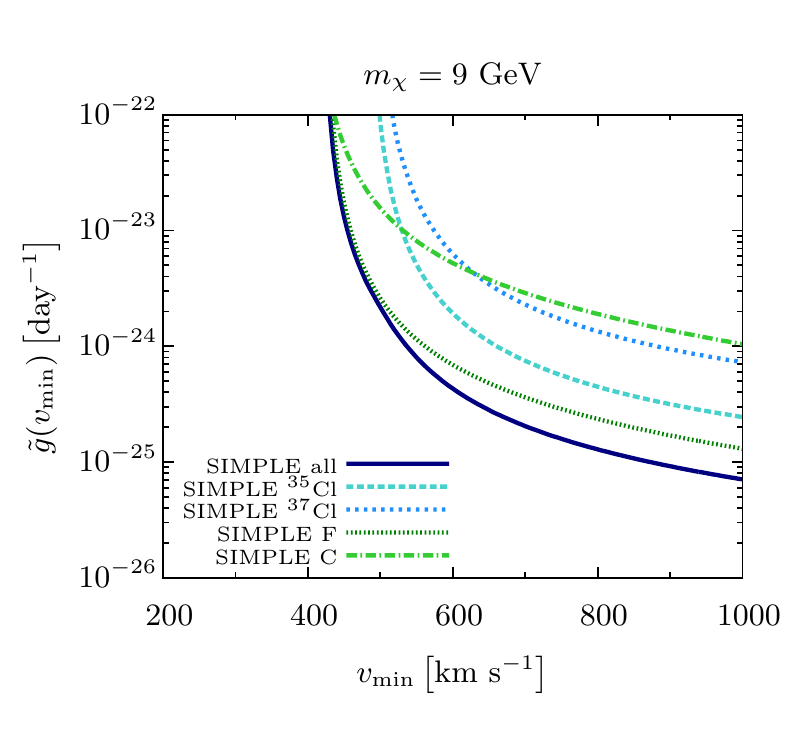} 
\caption{The individual bounds on $\tilde{g}(v_\text{min})$ for the different elements and isotopes as well as the combined total bound from the SIMPLE experiment.}
\label{fig:SIMPLE}
\end {figure}

For each element we then obtain a bound $\tilde{g}_{(i)}^\text{max}(v_\text{min})$
on the rescaled velocity integral. We can combine these bounds into a
total bound according to
\begin{equation}
\frac{1}{\tilde{g}^\text{max}(v_\text{min})} = \sum_i
\frac{1}{\tilde{g}_{i}^\text{max}(v_\text{min})} \; .
\end{equation}
This procedure is illustrated in Figure~\ref{fig:SIMPLE}.

In fact, it is sufficient to calculate $\tilde{g}_{i}^\text{max}(v_\text{min})$ for
only one of the elements in the target, as we can use this bound to
infer the bounds for all other elements. Suppose we have determined
the bound $\tilde{g}(v_\text{min}) <
\tilde{g}_1^\text{max}(v_\text{min})$ for $v_\text{min} = \hat{v}_1$
for an isotope with charge $Z_1$, mass number $A_1$ and form factor
$F_1$. For a different isotope, we can now infer the value of the
rescaled velocity integral at the minimal velocity $\hat{v}_2$, which
satisfies $E_1(\hat{v}_1) = E_2(\hat{v}_2)$, meaning that the two
velocities correspond to the same maximum recoil energy on the
respective nuclei. The bound from the second isotope is then given by
\begin{equation}
\tilde{g}_2^\text{max}(\hat{v}_2) = \frac{c_1 C^2_\mathrm{T}(Z_1,A_1)
  F_1(E_1(\hat{v}_1))^2}{c_2 C^2_\mathrm{T}(Z_2,A_2)
  F_2(E_2(\hat{v}_2))^2} \tilde{g}_1^\text{max}(\hat{v}_1) \; ,
\label{eq:isotopes}
\end{equation}
where $c_{1,2}$ are the respective concentrations of the isotopes.

\subsection{Measuring the velocity integral}
\label{ap:measure}

We turn now to the more difficult task of inferring information on
$\tilde{g}(v_\text{min})$ from a potential DM signal in an experiment
consisting of several different elements.  The obvious application is
the CRESST-II experiment, which we will use here for illustration. We
consider only the contribution of calcium and oxygen, because for DM
masses of $15$ GeV and less, nuclear recoils on tungsten have recoil
energies well below the threshold of the detector.

The differential event rate in CRESST-II is given by the sum of the
differential event rates for the individual elements, i.e.
\begin{equation}
  \left(\frac{\text{d}R}{\text{d}E_\text{R}}\right)^\text{tot} = 
\left(\frac{\text{d}R}{\text{d}E_\text{R}}\right)^{\text{(O)}} + 
\left(\frac{\text{d}R}{\text{d}E_\text{R}}\right)^{\text{(Ca)}} \; .
\label{eq:sum}
\end{equation}
To cause a recoil of energy $E_\text{R}$, a DM particle must have a
minimum velocity of $v_\text{min}^{\text{(O)}}(E_\text{R})$ for
scattering off oxygen and $v_\text{min}^{\text{(Ca)}}(E_\text{R})$ for
calcium. Consequently, if we measure the differential event rate at
the energy $E_\text{R}$, we simultaneously probe the rescaled velocity
integral at \emph{both} of these velocities. Therefore, we can only
extract information on
$\tilde{g}(v_\text{min}^{\text{(O)}}(E_\text{R}))$ if we know the
ratio of the differential event rates for scattering off oxygen and
calcium
\begin{equation}
\frac{\left(\text{d}R/\text{d}E_\text{R}\right)^{\text{(O)}}}{\left(\text{d}R/\text{d}E_\text{R}\right)^{\text{(Ca)}}}
= \frac{C^2_\mathrm{T}(Z^{\text{(O)}},A^{\text{(O)}}) \cdot
  F^{\text{(O)}}(E_\text{R})^2 \cdot
  g(v_\text{min}^{\text{(O)}}(E_\text{R}))}{C^2_\mathrm{T}(Z^{\text{(Ca)}},A^{\text{(Ca)}})
  \cdot F^{\text{(Ca)}}(E_\text{R})^2 \cdot
  g(v_\text{min}^{\text{(Ca)}}(E_\text{R}))} \; .
\label{eq:ratio}
\end{equation}

We present two different methods to determine this ratio. The first
makes use of the fact that the rescaled velocity integral is a
decreasing function of $v_\text{min}$ to place an upper bound on the
contribution from calcium. The second method requires a special
binning of the data in such a way that the contribution of calcium can
actually be determined from the combined information of several
bins. Once we have an estimate of the differential event rate from
oxygen alone, we can use the procedure described in
Section~\ref{sec:measureg} to obtain the corresponding values of
$\tilde{g}(v_\text{min})$.

\subsubsection*{Method 1}

This method makes use of the fact that $v_\text{min}^{\text{(Ca)}} >
v_\text{min}^{\text{(O)}}$ for the same recoil energy, because calcium
is the heavier nucleus and we are only interested in the case where $m_\chi < m_\text{O}$. We know that $\tilde{g}(v_\text{min})$ must
decrease as $v_\text{min}$ increases, so
$g(v_\text{min}^{\text{(Ca)}}(E_\text{R})) <
g(v_\text{min}^{\text{(O)}}(E_\text{R}))$. Substituting this
expression into Eq.~\eqref{eq:ratio}, it is possible to place a lower
bound on the contribution of oxygen
\begin{equation}
\left(\frac{\text{d}R}{\text{d} E_\text{R}}\right)^{\text{(O)}} >
\frac{C^2_\mathrm{T}(Z^{\text{(O)}},A^{\text{(O)}}) \cdot
  F^{\text{(O)}}(E_\text{R})^2}{C^2_\mathrm{T}(Z^{\text{(Ca)}},A^{\text{(Ca)}})
  \cdot F^{\text{(Ca)}}(E_\text{R})^2} \left(\frac{\text{d}R}{\text{d}
  E_\text{R}}\right)^{\text{(Ca)}} \equiv
S^{\text{(O,Ca)}}(E_\text{R}) \left(\frac{\text{d}R}{\text{d}
  E_\text{R}} \right)^{\text{(Ca)}}
\label{eq:calcium}
\end{equation}

If the differential event rate
$\left(\text{d}R/\text{d}E_\text{R}\right)^\text{tot}$ is known, we
can make use of this inequality and Eq.~\eqref{eq:sum} to infer that
\begin{equation}
\left(\frac{\text{d}R}{\text{d} E_\text{R}}\right)^{\text{(O)}} >
\left(\frac{\text{d}R}{\text{d} E_\text{R}}\right)^{\text{tot}}
\frac{S^{\text{(O,Ca)}}(E_\text{R})}{1 +
  S^{\text{(O,Ca)}}(E_\text{R})} \; .
\end{equation}
In practice, additional factors like the target composition and the
acceptance for each nuclear recoil need to be included in
$S^{\text{(O,Ca)}}(E_\text{R})$. In the case of CRESST-II, the minimum
contribution of oxygen in the relevant energy region is approximately
20\%.

Note, that this method only constrains, but does not actually
determine, the ratio of oxygen to calcium scatters in each bin. The
lower bound that we obtain is almost independent of energy although we
expect the oxygen fraction to approach 100\% as the recoil energy
increases. To reflect our ignorance of the true ratio, we do not show
central values for the measurements of $\tilde{g}(v_\text{min})$
obtained with this method. The second method, which we present below,
will allow for a more accurate determination of the ratio of oxygen to
calcium scatters.

\subsubsection*{Method 2}

In the approach discussed above, we considered each bin separately and
treated the presence of a second (heavier) element in the target as an
additional uncertainty which reduces the information that we can
extract from the data on the velocity integral. It would be desirable
to develop a method that combines the information from different bins
in order to use \emph{both} target elements to and constrain the
velocity integral more tightly. We now present such a method.

\begin{figure}[tb]
\centering
  \includegraphics[width=.5\columnwidth]{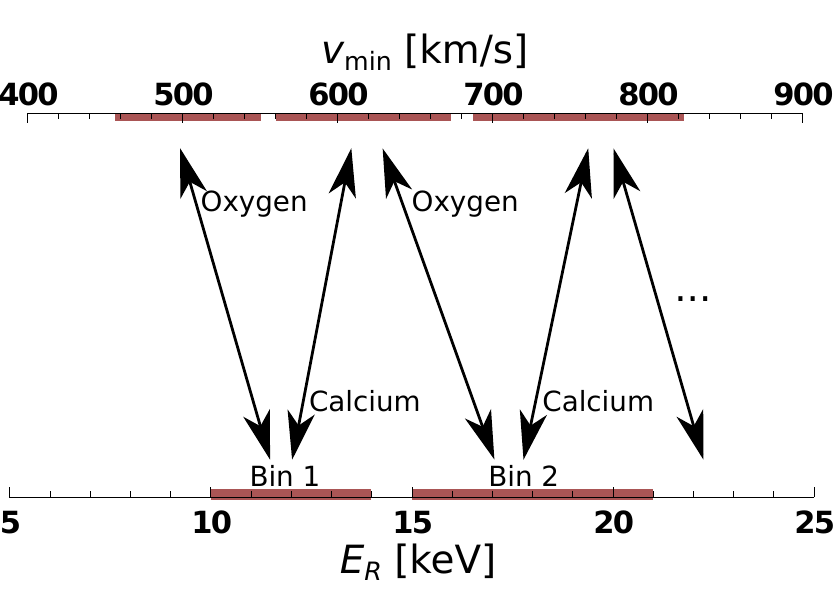}
\caption{An illustration of the method used to construct bins for CRESST-II. The two bins constructed in this way have the property that calcium recoils with energies in bin 1 probe the \emph{same} region of $v_\text{min}$-space as oxygen recoils with energies in bin 2. For this plot, we have assumed \mbox{$m_\chi = 9$ GeV}.}
\label{fig:method2}
\end {figure}

First of all, we must construct energy bins of different sizes
according to the following rule: Starting from an arbitrary bin
$\left[E^\text{min}_1, E^\text{max}_1\right]$, we use the heavier
target (assumed to be calcium) to convert the boundaries into
$v_\text{min}$-space and then the lighter target (oxygen) to convert
the velocities back into energy space. This way, we obtain a new
energy bin
$\left[E^\text{(O)}(v_\text{min}^\text{(Ca)}(E^\text{min}_1)),
  E^\text{(O)}(v_\text{min}^\text{(Ca)}(E^\text{max}_1))\right]$. The
two bins thus constructed have the property that calcium
recoils with energies in bin 1 probe the \emph{same} region of
$v_\text{min}$-space as oxygen recoils with energies in bin 2. We can
repeat this procedure recursively to obtain additional bins with
increasing bin size (see Figure~\ref{fig:method2}). It is always
possible to choose the first bin in such a way that the different bins
do not overlap.

We only wish to consider bins where a significant excess of signal
over expected background is observed, meaning that we abort the
iteration as soon as we obtain a bin where the signal is compatible
with the background estimate. To give a concrete example, for $m_\chi = 9$ GeV the energy
conversion factor between oxygen and calcium is
\begin{equation}
\frac{m_\text{Ca}}{\mu_\text{Ca}^2}\frac{\mu^2_\text{O}}{m_\text{O}}=1.50 \; ,
\end{equation}
so a possible construction for an experiment like CRESST-II would be \mbox{[8 keV, 12 keV]}, \mbox{[12 keV,
    18 keV]}, \mbox{[18 keV, 27 keV]}. Unfortunately, the current
CRESST-II data reaches down only to 10 keV and we cannot bin the data
arbitrarily so we will use only the two bins \mbox{[10 keV, 14 keV]}
and \mbox{[15 keV, 21 keV]} here. For different values of $m_\chi$ the conversion factor will also change, so different bins must be considered. We present possible choices for different DM masses in Table~\ref{tab:CRESSTbins}.

We start by evaluating the highest bin and then work our way
back towards bin~1. In the highest bin, we can reasonably assume that
the observed event rate is completely dominated by the lightest
element in the target. The reason is that we expect events in this bin
to arise from DM particles close to the escape velocity (otherwise,
there would also be events in higher energy bins). Consequently, in the
corresponding region of $v_\text{min}$-space, we expect the velocity
integral to decrease rapidly with increasing $v_\text{min}$
(see Figure~\ref{fig:HaloModels}) so that a lighter target element
should have a much larger event rate than a heavier one (probing the
velocity integral at higher velocities).

This assumption is good under two conditions. First, that the
different target nuclei are not too close in mass so that they
actually probe different regions of $v_\text{min}$-space and second,
that there is no additional source of background at large energies
that can hide a potential signal in higher energy bins. Both of these
conditions are fulfilled for the CRESST-II experiment, where the
background at large energies is approximately flat.

Under the assumption that only the lightest element contributes in the
highest bin, we can calculate $\tilde{g}(v_\text{min})$ for this bin
in the same way as described in Section~\ \ref{sec:measureg} for a target
consisting of a single element. Now we can exploit the fact that we
have constructed the bins in a special way and use the inferred value
of $\tilde{g}(v_\text{min})$ to \emph{predict} the number of calcium
events in the next lower bin.

\begin{table}[tb]
\centering
\begin{tabular}{c c c c c}
\hline\hline
$m_\chi$ / GeV & Conversion factor & Bin 1 / keV & Bin 2 / keV \\
\hline
6 & 1.72 & 10 -- 14 & 17 -- 24 \\
9 & 1.50 & 10 -- 14 & 15 -- 21 \\
12 & 1.34 & 12 -- 15 & 16 -- 20 \\
15 & 1.22 & & \\
\hline\hline
\end{tabular}
\caption{The energy conversion factors between oxygen recoils and calcium recoils for constant $v_\text{min}$ for different choices of the DM mass $m_\chi$. The third and fourth column give possible choices for two bins that are related by this conversion factor. For $m_\chi = 15$ GeV, a different binning of the CRESST-II data would be required to construct non-overlapping bins.}
\label{tab:CRESSTbins}
\end{table}

We can then subtract the predicted number of calcium events in the
next lower bin to obtain the number of events that is due to oxygen
scatters alone. Again, we are left with a number of events that
correspond to scattering of the lightest element alone. We can use
this number to calculate $\tilde{g}(v_\text{min})$ in the region of
$v_\text{min}$-space corresponding to the second to highest bin. Using
this measurement to predict the calcium scatters in the next lower
bin, we can recursively extract $\tilde{g}(v_\text{min})$ for all
bins. Compared to the previous method, the advantage is clearly a much
more accurate prediction of $\tilde{g}(v_\text{min})$. The
disadvantage is that we are forced to bin the data in a certain way,
which is inconvenient if the energy of each event is not publicly
available.

\section{Constraining the modulation fraction using the absolute rate}
\label{ap:modfrac}

We will demonstrate that it is possible to constrain the modulation
fraction once we have determined the average value of the rescaled
velocity integral from a direct detection experiment, without assuming
any specific properties of the DM halo. The basic idea is that we can
place an upper bound on the maximum value, $\tilde{g}(v_\text{min},
\text{ summer})$, and a lower bound on the minimum value,
$\tilde{g}(v_\text{min}, \text{ winter})$, using only the time
average, $\tilde{g}(v_\text{min})$.

\begin{figure}[tb]
\begin{center}
{
\includegraphics[width=.99\columnwidth]{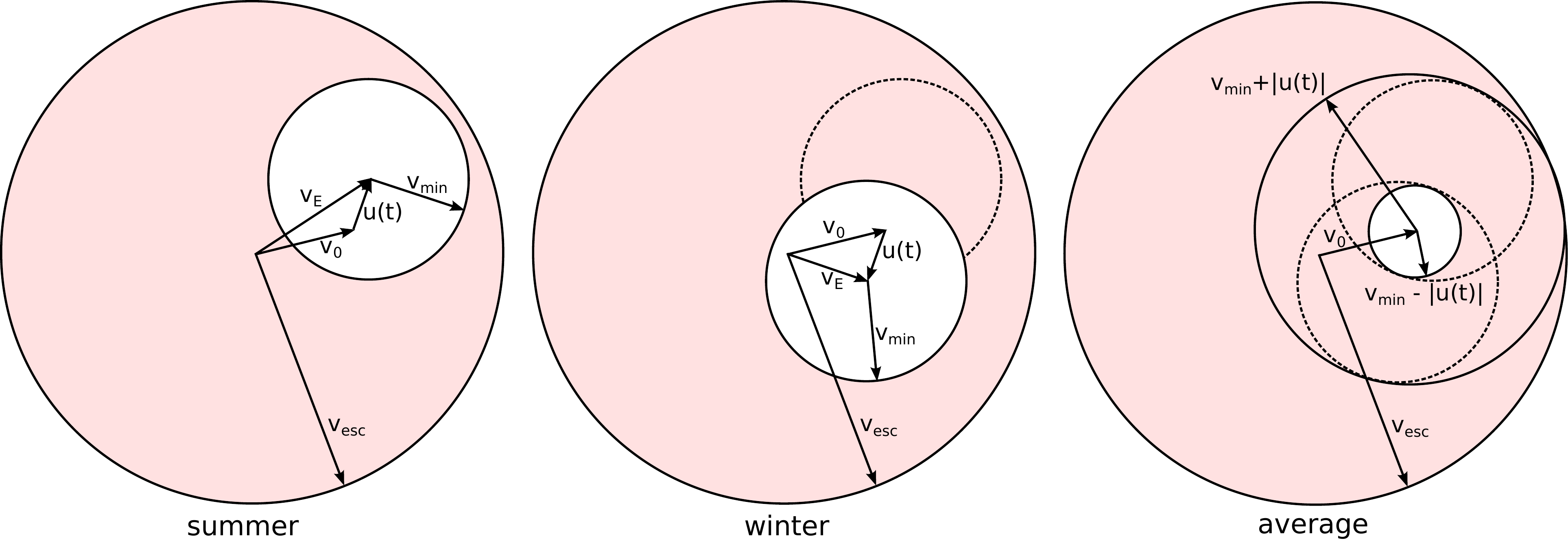}
}
\caption{A graphical representation of the velocity integral. The red shaded region is the region of velocity space that is being integrated over, while the white regions are excluded by the requirement $v > v_\text{min}$. We can place an upper bound on the maximum value, $\tilde{g}(v_\text{min}, \text{ summer})$, and a lower bound on the minimum value, $\tilde{g}(v_\text{min}, \text{ winter})$, using only the time average, $\tilde{g}(v_\text{min})$. Note that the relative magnitude of the different velocities are not drawn to scale.}
\label{fig:vspace}
\end{center}
\end {figure}

How such bounds can be obtained is sketched in
Figure~\ref{fig:vspace}. We observe that
\begin{align}
\tilde{g}(v_\text{min},\text{ summer}) & < \tilde{g}(v_\text{min} - u)
\\ \tilde{g}(v_\text{min},\text{ winter}) &> \tilde{g}(v_\text{min} +
u)
\end{align}
where $u = 29.8$ km/s is the velocity of the Earth around
the Sun. From this it follows that
\begin{equation}
\Delta \tilde{g}(v_\text{min}) = \tilde{g}(v_\text{min},\text{
  summer}) - \tilde{g}(v_\text{min},\text{ winter}) <
\tilde{g}(v_\text{min} - u) - \tilde{g}(v_\text{min} + u) ,
\end{equation}
and therefore, that the modulation fraction is bounded by 
\begin{equation}
A(v_\text{min}) = \frac{\tilde{g}(v_\text{min},\text{ summer}) -
  \tilde{g}(v_\text{min},\text{
    winter})}{\tilde{g}(v_\text{min},\text{ summer}) +
  \tilde{g}(v_\text{min},\text{ winter})} <
\frac{\tilde{g}(v_\text{min} - u) - \tilde{g}(v_\text{min} + u)}{2
  \tilde{g}(v_\text{min})} ,
\end{equation}
where we have used $\tilde{g}(v_\text{min},\text{ summer}) +
\tilde{g}(v_\text{min},\text{ winter})\approx 2 
\tilde{g}(v_\text{min})$.

Note that an even stronger constraint can be obtained under the
assumption of an isotropic DM halo. In this case, only the seasonal
change in the \emph{radial} component of $\boldsymbol{u}$, $u_r$,
matters, because the halo will look the same for any value of $u_\phi$
and $u_\theta$. In this case, we can replace $u$ by $u_r = u \cos
\gamma = 15.0$ km/s in the equation above, where $\gamma$ is the
inclination angle between the galactic and ecliptic plane.

\section{Various descriptions of the DM halo}

\label{ap:halos}

We present the velocity distributions considered in
Sections~\ref{sec:results} and \ref{sec:modulations} and provide the
parameter choices used for Figures~\ref{fig:HaloModels} and
\ref{fig:ModulationConstraints} --- for further details see the 
references. First, we discuss isotropic velocity distributions, where
$f(\boldsymbol{v})$ depends only on $|\boldsymbol{v}|$ and there
is no preferred direction for DM velocities at the position
of the Earth. Then we  consider models where the anisotropy parameter
\begin{equation}
\beta(r) \equiv 1 - \frac{\langle v_\theta^2 \rangle + \langle
  v_\phi^2 \rangle}{2\langle v_r^2 \rangle}
\end{equation}
is non-zero, motivated by numerical simulations which favour radially
biased orbits with $\beta = 0.1-0.4$ at the position of the Earth.

\subsection{Isotropic models}

\subsubsection*{Standard Halo Model}

In the Standard Halo Model (SHM), the velocity distribution function
is given by
\begin{equation}
f(\boldsymbol{v}) = \left\{
     \begin{array}{lr}
       \frac{1}{N} \left[ \exp (-v^2 / v_0^2) - \exp (-v_\text{esc}^2
         / v_0^2) \right] \ \ & |\boldsymbol{v}| < v_\text{esc}\\ 0 &
       |\boldsymbol{v}| \geq v_\text{esc}
     \end{array}
   \right. \; ,
\end{equation}
where $N$ is a normalisation constant. Possible values for the
velocity dispersion $v_0$ range from 180 km/s to 280 km/s, while
$v_\text{esc}$ can vary between 450 km/s and 650 km/s (see also the
left panel of Figure~\ref{fig:HaloModels}). Unless explicitly stated
otherwise, we will always assume the standard values $v_0 = 220$ km/s
and $v_\text{esc} = 544$ km/s.

\subsubsection*{King models}

The cut-off at $v_\text{esc}$ is introduced by hand in the SHM, which
would otherwise predict particles with infinitely high velocities and
an infinite mass for the Galaxy. This problem is addressed in the King
model, which provides a cut-off in a self-consistent manner. The
velocity distribution is obtained from the distribution function:
\begin{equation}
f(\mathcal{E}) = \left\{
     \begin{array}{lr}
       \frac{1}{N} \left[ \exp \left(\mathcal{E}/\sigma^2\right) - 1
         \right] , \ \ & \mathcal{E} > 0\\ 0 , & \mathcal{E} \leq 0
     \end{array}
   \right. \; ,
\end{equation}
where $\mathcal{E} = \Psi(\boldsymbol{x}) - v^2/2$ and
$\Psi(\boldsymbol{x})$ is the relative gravitational potential. The
local escape velocity at a position $\boldsymbol{x}$ is given by
$v_\text{esc} = \sqrt{2\Psi(\boldsymbol{x})}$. At the position of the
Earth $f(\boldsymbol{v})$ can be parameterised in the same way as the SHM but the
parameter $\sigma$ is not directly linked to the velocity dispersion
and can therefore take values that are much larger than in the SHM
\cite{Chaudhury:2010hj, Kundu:2011ek}. Nevertheless, due to its
similarity to the SHM we will not discuss King models further.

\subsubsection*{Double power-law profiles}

A simple modification of the SHM was introduced in
Ref.\cite{Lisanti:2010qx}. For double power-law density profiles such
as the NFW-profile \cite{Navarro:1995iw}, the following ansatz for the
velocity distribution reproduces better the behaviour at high
velocities:
\begin{equation}
f(\boldsymbol{v}) = \left\{
     \begin{array}{lr}
       \frac{1}{N} \left[ \exp \left(\frac{v_\text{esc}^2-v^2}{k
           v_0^2}\right) - 1 \right]^k ,\ \ & |\boldsymbol{v}| <
       v_\text{esc}\\ 0 ,& |\boldsymbol{v}| \geq v_\text{esc}
     \end{array}
   \right. \; .
\end{equation}
Setting the power-law index $k$ equal to 1 recovers the SHM. The
choice $1.5 \leq k \leq 3.5$ is found to give a better fit to velocity
distributions extracted from $N$-body simulations. We use $k = 2.5$
throughout.

\subsubsection*{Tsallis model}

It was argued \cite{Ling:2009eh} that 
the velocity distribution of dark matter particles in numerical simulations including baryons can be well described by
\begin{equation}
f(\boldsymbol{v}) = \left\{
     \begin{array}{lr} \frac{1}{N} \left[ 1 - \left(1-q\right)\frac{\boldsymbol{v}^2}{v_0^2}\right]^{1/(1-q)} ,\ \ & |\boldsymbol{v}| < v_\text{esc}\\
       0 ,& |\boldsymbol{v}| \geq v_\text{esc}
     \end{array}
   \right. \; ,
\end{equation}
see also Ref.\cite{Vergados:2007nc}. We adopt the parameters $q=0.773$, $v_0 = 267.2$ km/s and $v_\text{esc} = 560.8$ km/s from Ref.\cite{Ling:2009eh}.

\subsection{Anisotropic models}

\subsubsection*{Numerical simulations}

A simple anisotropic model has been proposed \cite{Kuhlen:2009vh} to
describe the data from numerical $N$-body simulations such as Via
Lactea \cite{Diemand:2006ik, Diemand:2008in}, GHALO
\cite{Stadel:2008pn} or Aquarius \cite{Springel:2008cc}:
\begin{equation}
f(\boldsymbol{v}) = \left\{
     \begin{array}{lr}
       \frac{1}{N} \left[ \exp \left(-(v_r^2 /
         \overline{v}_r^2)^{\alpha_r}\right) \exp \left(-(v_t^2 /
         \overline{v}_t^2)^{\alpha_t}\right) \right] ,\ \ &
       |\boldsymbol{v}| < v_\text{esc}\\ 0 ,& |\boldsymbol{v}| \geq
       v_\text{esc}
     \end{array}
   \right. \; ,
   \label{eq:numsim}
\end{equation}
where $v_t = \sqrt{v_\theta^2 + v_\phi^2}$. For the figures we take
the best-fit parameters for the Via Lactea II simulation, namely $v_r
= 202.4$ km/s, $v_t = 128.9$ km/s, $\alpha_r = 0.934$ and $\alpha_t =
0.642$ \cite{Kuhlen:2009vh}, but we also show the velocity integral
and modulation fraction observed in the GHALO$_{\text{s}}$ simulation.

\subsubsection*{Logarithmic ellipsoidal model}

The simplest triaxial generalisation of the velocity distributions
considered above was discussed in Refs.\cite{Evans:2000gr,
  Green:2002ht, Belli:2002yt}.  We allow a different velocity
dispersion in all three directions, giving
\begin{equation}
f(\boldsymbol{v}) = \left\{
     \begin{array}{lr}
       \frac{1}{N} \left[ \exp \left(-v_r^2 /
         \overline{v}_r^2-v_\phi^2 / \overline{v}_{\phi}^2-v_z^2 /
         \overline{v}_z^2\right) \right] ,\ \ & |\boldsymbol{v}| <
       v_\text{esc}\\ 0 ,& |\boldsymbol{v}| \geq v_\text{esc}
     \end{array}
   \right. \; .
\end{equation}
The three parameters $\overline{v}_r$, $\overline{v}_\phi$ and
$\overline{v}_z$ depend on two constants $p$ and $q$ that describe the
density distribution and the isotropy parameter $\gamma$ (as well as
$v_0$). Following Ref.\cite{Green:2002ht}, we take $p = 0.9$, $q =
0.8$ and $\gamma = -1.33$ and calculate $\overline{v}_r$,
$\overline{v}_\phi$ and $\overline{v}_z$ under the assumption that the
Earth is on the major axis.

\subsubsection*{Distribution functions with $\beta = 0.5$}

For a constant anisotropy of $\beta = 0.5$, it is possible to
calculate the velocity distribution from a given density profile. This
was done \cite{Evans:2005tn}, for centrally cusped density profiles of
the form
\begin{equation}
\rho \propto \frac{a^{b-2}}{r(r+a)^{b-1}} \; .
\end{equation}
For example, for $b=4$, corresponding to the Hernquist density profile
\cite{Hernquist:1990be}, the velocity profile is given by
\begin{equation}
f(\boldsymbol{v}) = \left\{
      \begin{array}{lr}
        \frac{1}{N} \frac{1}{v_\text{esc} \sqrt{v_x^2 + v_y^2}} \left(
        1 - \frac{v_x^2 + v_y^2 + v_z^2}{v_\text{esc}^2} \right)^2 ,
        \ \ & |\boldsymbol{v}| < v_\text{esc}\\ 0 ,& |\boldsymbol{v}|
        \geq v_\text{esc}
      \end{array}
    \right. \; .
\end{equation}
For $b=3$, corresponding to the NFW profile \cite{Navarro:1995iw}, an
analytical expression of the velocity distribution does not exist, but
it is straightforward to numerically calculate the velocity
distribution from the density profile.  For the plots shown we adopt
$a = 10$ kpc.

\section{Overview of direct detection experiments}

\label{ap:experiments}

We will briefly discuss the direct detection experiments that we have
considered in this paper and state the assumptions that we have made.
To derive constraints from XENON100, XENON10, CDMS-Ge and CRESST-II,
we employ the `maximum gap' method \cite{Yellin:2002xd}, while for the
low threshold analysis of CDMS-Ge and CDMS-Si, we use the `binned
Poisson' method \cite{Green:2001xy, Savage:2008er}. Our best-fit regions are
calculated using a $\chi^2$ parameter estimation method.\footnote{For
  CRESST-II we use Eq.~(13) from Ref.\cite{Fairbairn:2008gz} to
  calculate the best-fit parameter region.}

\subsection*{XENON100}

We use the most recent data from 100.9 live days of data taking
\cite{Aprile:2011hi}. For the relative scintillation efficiency
$\mathcal{L}_\text{eff}$ we use a logarithmical extrapolation to zero
and calculate the energy resolution under the assumption that it is
dominated by Poisson fluctuations in the number of photoelectrons.

\subsection*{XENON10}

It is important to note that the published XENON10 data from Ref.\cite{Angle:2007uj} was based on outdated assumptions concerning
the relative scintillation efficiency, so all energy scales must be
rescaled to the same $\mathcal{L}_\text{eff}$ as for the XENON100
data. We calculate the energy resolution in the same way as for
XENON100.

We also consider the $S2$-only analysis presented in
Ref.\cite{Angle:2011th}. We adopt their conservative assumption for
the ionisation yield with a sharp cut-off at 1.4~keV and take a flat
detector acceptance of 0.41. To calculate the energy resolution we
assume that the production of electrons is governed by Poisson
statistics. We also apply the $z$-cut suggested by the XENON10
collaboration to reduce the number of observed events.

\subsection*{CoGeNT}

We analyse the publicly available data from the CoGeNT experiment over
the course of 1.2 years \cite{Aalseth:2011wp}. To determine the total
event rate, we subtract the L-shell EC contribution and a constant
background. We use the Lindhard quenching factor and the (corrected)
detector resolution from Ref.\cite{Aalseth:2008rx}. The efficiency
curve was provided by J. Collar (private communication). Our bin width
is 0.05 keVee.

To determine the modulation, we fix the peak date to $t_0 = 146$ days,
which is the best-fit value for the DAMA modulation. The bins
considered and the resulting event rates and modulation amplitudes are
given in Table~\ref{tab:Cogent}.

\begin{table}
\centering
\begin{tabular}{c c c c c}
\hline\hline
Bin / keVee & Event rate & Error \\
& in cpd kg$^{-1}$ keVee$^{-1}$ & in cpd kg$^{-1}$ keVee$^{-1}$ \\
\hline
0.50 -- 0.61 & 12.8 & 1.0 \\
0.61 -- 0.72 & 10.1 & 0.9 \\
0.72	-- 0.85 & 6.7 & 0.8 \\
0.85 -- 1.00 & 5.5 & 0.8 \\
1.00 -- 1.15 & 4.5 & 0.7 \\
\hline
Bin / keVee & Modulation amplitude & Error \\
& in cpd kg$^{-1}$ keVee$^{-1}$ & in cpd kg$^{-1}$ keVee$^{-1}$ \\
\hline
0.5 -- 1.0 & 0.75 & 0.54 \\
1.0 -- 2.0 & 0.51 & 0.30 \\
2.0 -- 3.2 & 0.26 & 0.16\\
\hline\hline
\end{tabular}
\caption{The CoGeNT event rate and its modulation: the modulation
  amplitude has been obtained by fitting a cosine with period fixed to
  365 days and the phase fixed to $t_0 = 146$ days.}
\label{tab:Cogent}
\end{table}

\subsection*{CDMS-Ge}

We use the final data \cite{Ahmed:2009zw} corresponding to an exposure
of 612 kg days. We assume a detector resolution of 0.4 keV and take
the detector acceptance from Ref.\cite{Ahmed:2008eu}.

We also consider the dedicated low-threshold analysis of the CDMS-Ge
data \cite{Ahmed:2010wy} and use only the data of the most
constraining detector, T1Z5. Here, we adopt an energy resolution of
$0.2\,(E_\text{R}\cdot\mathrm{keV})^{1/2}$
\cite{SchmidtHoberg:2009gn}.

\subsection*{CDMS-Si}

To obtain a constraint from the silicon detectors in CDMS-II, we take
the low-threshold data from the shallow-site run
\cite{Akerib:2010pv}. We include the runs at 3V and at 6V but consider
only the Z4 detector which gives the strongest bound. Resolution and
detector efficiency are both taken from Ref.\cite{Akerib:2010pv}.

\subsection*{SIMPLE}

We do not attempt to combine the results from Stage 1
\cite{Felizardo:2010mi} and Stage 2 \cite{Felizardo:2011uw} but
consider only the latter. Since the recoil energy is not measured, the
experiment can place only an upper limit on the total event rate above
the detector threshold of 8 keV.  We take the cut acceptance and
nucleation efficiency from Ref.\cite{Felizardo:2011uw}.

\begin{table}
\centering
\begin{tabular}{c c c c c}
\hline\hline
Bin / keV & Total events & Expected background & Signal (min) & Signal (max) \\
\hline
10 -- 13 & 9 & 3.2 & 3.2 & 9.6 \\
13 -- 16 & 15 & 6.1 & 5.3 & 13.3 \\
16 -- 19 & 11 & 7.0 & 1.2 & 7.8 \\
19 -- 22 & 8 & 6.3 & 0.2 & 5.0 \\
22 -- 25 & 4 & 5.2 & 0 & 1.8 \\
25 -- 28 & 4 & 4.6 & 0 & 2.2 \\
28 -- 31 & 4 & 4.3 & 0 & 2.5 \\
31 -- 34 & 3 & 4.0 & 0 & 1.5 \\
34 -- 37 & 4 & 3.7 & 0 & 3.0 \\
37 -- 40 & 5 & 3.5 & 0.2 & 4.3 \\
\hline
13 -- 15 & 11 & 3.8 & 4.0 & 11.0 \\
16 -- 19 & 11 & 7.0 & 1.2 & 7.8 \\
19 -- 22 & 8 & 6.3 & 0.2 & 5.0 \\
\hline
10 -- 14 & 18 & 5.4 & 8.4 & 17.4 \\
12 -- 15 & 19 & 5.7 & 9.1 & 18.1 \\
15 -- 21 & 19 & 13.7 & 1.7 & 10.1 \\
16 -- 20 & 13 & 9.4 & 0.9 & 7.9 \\
17 -- 24 & 17 & 14.1 & 0.5 & 7.7 \\
\hline\hline
\end{tabular}
\caption{Data from the CRESST-II experiment: total number of events,
  expected number of background events and $1\sigma$ confidence
  interval for the signal using a Feldman-Cousins approach.}
\label{tab:CRESSTII}
\end{table}

\subsection*{CRESST-II}

We use the recent results from 730 kg days of data taking
\cite{Angloher:2011uu}. We estimate that the acceptance region
corresponds to $86\%$ acceptance for oxygen, $89\%$ for tungsten and
$90\%$ for calcium. We take the energy resolution to be $0.3$ keV for
each detector. To extract the DM signal, we read off the total number
of measured events and the expected background from Figure~11 of
Ref.\cite{Angloher:2011uu} and use the Feldman-Cousins approach
\cite{1998PhRvD573873F}.

Depending on the application, we bin the CRESST-II data in different
ways. In order to calculate best-fit parameter regions, we consider a
bin width of 3 keV, rather than a bin width of 1 keV as presented in
Ref.\cite{Angloher:2011uu}. However, if we want to calculate
$\tilde{g}(v_\text{min})$ using Method 1 from
Appendix~\ref{ap:measure}, we have to make sure that the detector
efficiency does not vary strongly within a single bin due to the onset
of additional detector modules. This condition is fulfilled for the
three bins \mbox{[13 keV -- 15 keV]}, \mbox{[16 keV -- 19 keV]} and
\mbox{[19 keV -- 22 keV]}. Finally, to apply Method 2 from
Appendix~\ref{ap:measure}, we need to bin the data in a special way,
namely in the two bins \mbox{[10 keV -- 14 keV]} and \mbox{[15 keV --
    21 keV]}. The resulting number of events is shown in
Table~\ref{tab:CRESSTII}.

Finally, we also consider the improved constraint from the CRESST-II
commissioning run presented in Ref.\cite{Brown:2011dp} using the
detector efficiencies given there.

\subsection*{DAMA}

We use the combined observed event rates from DAMA/NaI and DAMA/LIBRA
\cite{Bernabei:2010mq} for the 8 bins spanning the energy range
2-6~keVee. No significant modulation is observed at higher
energies. As we are interested in DM particles with mass below 20~GeV,
we neglect scattering on iodine and consider only the contribution of
sodium. We assume a quenching factor of 0.3 (other values are
discussed in Refs.\cite{Kopp:2011yr, Kelso:2011gd}) and no
contribution from channelling \cite{Bozorgnia:2010xy}.

\end{appendix}

\providecommand{\href}[2]{#2}\begingroup\raggedright\endgroup

\end{document}